\documentclass[aps,twocolumn,preprintnumbers,nofootinbib,prd,superscriptaddress,10pt]{revtex4-1}

\usepackage[utf8]{inputenc}

\usepackage{epsf}  
\usepackage{graphicx}
\usepackage{amsmath}
\usepackage{amsfonts,color}
\usepackage{amssymb,float}

\begin{document}
\title{Black-Hole Perturbation Plus Post-Newtonian Theory: Hybrid Waveform for Neutron Star Binaries}

\author{Xuefeng Feng}
\email{fengxuefeng@amss.ac.cn}
\affiliation{Institute of Applied Mathematics, Academy of Mathematics and Systems Science, Chinese Academy of Science, Beijing 100190, China}
\affiliation{School of Mathematical Sciences, University of Chinese Academy of Sciences, Beijing 100049, China}
\affiliation{University of Guelph, Guelph, Ontario N1G 2W1, Canada}

\author{Zhenwei Lyu}
\affiliation{University of Guelph, Guelph, Ontario N1G 2W1, Canada}
\affiliation{Perimeter Institute for Theoretical Physics, Waterloo, Ontario N2L 2Y5, Canada}

\author{Huan Yang}
\email{hyang@perimeterinstitute.ca}
\affiliation{University of Guelph, Guelph, Ontario N1G 2W1, Canada}
\affiliation{Perimeter Institute for Theoretical Physics, Waterloo, Ontario N2L 2Y5, Canada}

\begin{abstract}
We consider the motion of nonspinning, compact objects orbiting around a Kerr black hole with tidal couplings. The tide-induced quadrupole moment modifies both the orbital energy and outgoing fluxes, so that over the inspiral timescale there is an accumulative shift in the orbital and gravitational wave phase. Previous studies on compact object tidal effects have been carried out in the Post-Newtonian (PN) and Effective-One-Body (EOB) formalisms.  In this work, within the black hole perturbation framework, we propose to characterize the tidal influence in the expansion of mass ratios, while higher-order PN corrections are naturally included. For the equatorial and circular orbit,  we derive the leading order, frequency dependent tidal phase shift which agrees with the  Post-Newtonian result at low frequencies but deviates at high frequencies. We also find that such phase shift has weak dependence ($\le 10\%$) on the spin of the primary black hole. Combining this black hole perturbation waveform with the Post-Newtonian waveform, we propose a frequency-domain, hybrid waveform that shows comparable accuracy as the EOB waveform in characterizing the tidal effects, as calibrated by numerical relativity simulations. Further improvement is expected as the next-to-leading order in mass ratio and the higher-PN tidal  corrections are included. This hybrid approach is also applicable for generating binary black hole waveforms.
\end{abstract}

\maketitle

\section{Introduction}
Inspiraling and coalescing  compact-object binary systems, including black holes and/or neutron stars, are important sources of ground-based gravitational waves (GW) detectors, e.g. LIGO\cite{gonzalez2004search} and Virgo\cite{acernese2004search}. Up to the O3 observation run, Advanced LIGO and Virgo have detected more than thirty binary black hole mergers, two binary neutron star mergers and one possible black hole-neutron star merger. The number of events is expected to increase significantly as Advanced LIGO and Virgo reach their design sensitivities.

Constructing GW waveform models are crucial for efficiently detecting  these binary systems as well as accurately estimating their source properties based on the observation data. Since it is computationally expensive to numerically solve Einstein's equation (and associated hydrodynamical equations if a neutron star is involved) for the binary evolution across the entire observation frequency band, especially with the large parameter space needed to characterize these binaries, several (semi)-analytical or phenomenological methods have been developed to complement the information from numerical simulations and generate reliable waveforms \cite{ajith2011inspiral,pan2014inspiral,field2014fast,van2020intermediate}. 

These methods generally follow  different avenues of analytical approximations in modelling the binary black hole inspiral waveform. For example, the low-frequency inspiral dynamics and associated waveform are treated within the Post-Newtonian (PN) framework in the ``Phenom" waveform series \cite{ajith2011inspiral,khan2016frequency}. At higher frequency certain calibrations with numerical waveforms are performed to bridge the gap between the PN inspiral description with the black hole ringdown. On the other hand, the PN expansion is restructured in the Effective-One-Body formalism \cite{buonanno1999effective} through a mapping to an effective spacetime of the relative motion, so that the resumed PN results may be better attached to the strong-gravity regime. Calibration with numerical relativity data has also been used to improve the accuracy of EOB waveforms.



When the mass ratio between the secondary and the primary black hole is small, we can view the smaller black hole as a particle moving in a perturbed spacetime of the primary black hole, where the metric perturbation and associated dynamical effects can be evaluated in a systematic expansion in the mass ratio. 
This black-hole-perturbation approach is the leading solution to produce waveforms of extreme mass-ratio inspirals (EMRIs), which are important sources for space-borne GW detectors such as  LISA \cite{amaro2017laser}. Given this expansion scheme, it is then natural to ask what is  its regime of applicability in mass ratios?  Interestingly, recent studies \cite{le2011periastron,favata2004black,le2013periastron,le2014overlap,zimmerman2016redshift,van2017self,le2018horizon,rifat2020surrogate,van2020intermediate,anninos1995head,fitchett1984linear,sperhake2011collisions,le2012gravitational,nagar2013gravitational} on this question have revealed a rather surprising result: the EMRI-based waveform may be even applicable for equal-mass binaries. 
In particular, for the equatorial and circular orbit, the GW phase can be written as the post-adiabatic expansion\cite{van2020intermediate}
\begin{align}\label{eq:phaseexp}
\psi(\omega)= \frac{\psi_{0}(\omega)}{\eta} +\psi_1(\omega) + \eta \psi_2(\omega) +...\,,
\end{align}
where $\omega$ is the orbital angular frequency and $\eta = m_1 m_2/(m_1+m_2)^2$ is the symmetric mass ratio, the function $\psi_i(\omega)$ is the coefficient of the order $\eta^{i}$ term. When the mass ratio is extreme, the symmetric mass ratio is almost the same as the mass ratio $q=m_1/m_2\leq 1$. The comparison with numerical relativity waveforms shows that, across the entire inspiral frequency range,  high order terms (starting from $\psi_2$ in the expansion) only contribute $\le 2$ radians phase shift even for equal-mass black hole binaries (with $\eta=1/4$) for most of the frequency range, except near the transition regime from inspiral to plunge \footnote{It is expected that an additional correction of order $\eta^{-1/5}$ must be introduced to account for the transition effects \cite{buonanno2000transition,ori2000transition}.}. This observation indicates that Eq.~\eqref{eq:phaseexp} may be a fast-converging series even for equal-mass binaries, so that the first several terms may suffice to produce accurate waveforms.

If at least one of the compact objects in the binary is a neutron star, tide-induced neutrons star deformation has to be included into the binary dynamics. This effect was first computed in \cite{flanagan2008constraining} for the leading order term in the waveform, with higher order PN corrections worked out in \cite{vines2011post}. Later on these PN tidal corrections were incorporated in the EOB framework, for both the equilibrium tide \cite{nagar2018time} and the dynamic tide \cite{hinderer2016effects}.


In this work, we adopt the black hole perturbation point of view, and evaluate the induced quadrupole moment of a neutron star moving in a perturbed spacetime of the primary black hole. In the local rest frame (or more precisely, within the ``asymptotically Cartesian and mass centered" coordinates \cite{thorne1998tidal} \footnote{In the multipole expansion picture discussed in \cite{thorne1980multipole}, the central object can be fully relativistic. As the multipole moments are derived in the asymptotic zone, Eq.~\eqref{eq:qe} can be viewed as the definition for the relativistic Love number $\lambda$.}) of the neutron star and in adiabatic approximation, the induced quadrupole moment is
\begin{eqnarray} \label{eq:qe}
	Q_{ab}=-\lambda E_{ab}
\end{eqnarray}
where $E_{ab}$ is the tidal tensor in the local spacetime and $\lambda$ is  the tidal Love number. In the equilibrium tide approximation, $\lambda$ is assumed to be a constant; with dynamical tide included, $\lambda$ can be thought as a function of the orbital frequency. Additional subtlety comes in if the orbital evolution  cross one or more mode resonances, where residual free mode oscillations will be present after these resonances and Eq.~\eqref{eq:qe} breaks down \cite{yang2019inspiralling}. For the purpose of this study, since the primary mode (f-mode) generally has frequency higher than the inspiral frequency, we will assume that the adiabatic approximation holds in the entire inspiral frequency range. Discussions on mode resonances and their detectability with LIGO and future detectors can be found in \cite{pan2020probing,yang2018evolution,yang2019inspiralling,poisson2020gravitomagnetic,schmidt2019frequency}.

In the black hole perturbation picture, the metric perturbation generated by the less massive black hole can be expanded in power laws of the mass ratio $h =h_1 q + h_2 q^2+ ...$, with $q = m_1/m_2$, and the less massive black hole $m_1$ can be viewed as moving along geodesics of the spacetime with metric $g_{\rm Kerr} + h$ \cite{detweiler2001radiation}. This mass ratio expansion of $h$ justifies the mass ratio expansion of $\phi$ in Eq.~\eqref{eq:phaseexp}.
When the less massive object is a neutron star, its motion can be viewed as a perturbed geodesic of the spacetime $g_{\rm Kerr}+ h$. This deviation from geodesic mainly comes from multipole interaction between the star and its environmental tidal field, while $h$ is sourced by the monopole (``the point-mass" piece), quadrupole, and all higher order multipole  parts of the stress-energy tensor. For simplicity, we truncate the multipole expansion at the quadrupole order and use the Mathisson-Papapetrou-Dixon prescription \cite{dixon1964covariant} to construct the stress-energy tensor of the star. 
To the linear order in $\lambda$, the tidal energy of the object and the tidal induced gravitational radiation flux are all $1/q$ or $1/\eta$ order lower than those of a point mass, so that the correction to the gravitational phase starts at $q^{-2}$ or $\eta^{-2}$ order.
Both $q$ and $\eta$ are  eligible choices of expansion parameters  in the small mass ratio limit, but they will give rise to rather different result as we truncate the series and apply it in the comparable mass ratio limit. For binary black hole waveforms it seems $\eta$ is a more efficient expansion parameter \cite{van2020intermediate}, but for tidal corrections the optimal choice is yet to be determined.

The leading-PN-order tidal correction to the gravitational wave phase is 
\begin{align}
\delta \psi = -\frac{9}{16} \frac{v^5}{\mu M^4} \left [ \left ( \frac{11m_2}{m_1}+\frac{M}{m_1} \right ) \lambda_1+1 \leftrightarrow 2\right ]
\end{align}
with $\mu$ being the reduced mass $m_1 m_2/M =\eta M$ and $M$ is the total mass $M=m_1+m_2$.
This motivates us to write down the tide-induced phase shift contributed by the less massive star (star ``1") as
\begin{align}\label{eq:tideA}
\eta \psi_{\rm BP,1} \approx \lambda_{\rm 1} (q^{-1}\psi^{(-1)}_{\rm BP} + \psi^{(0)}_{\rm BP}  + \sum_{n \ge 1} \eta^n  \psi^{(n)}_{\rm BP})\,.
\end{align}
which naturally includes all PN corrections, with the subscript ``BP" denoting ``Black Hole Perturbation". In particular, the $\psi^{(-1)}_{\rm BP}$ term can be obtained considering the tidal deformation of the neutron star due to the background Kerr spacetime of the primary black hole, and $\psi^{(0)}_{\rm BP}$ corresponds to the extra tidal deformation induced by $h_1$. 
If the companion is also a neutron star, its tidal contribution to the waveform can be obtained by replacing $q$ by $q^{-1}$, $\lambda_{\rm 1}$ by $\lambda_{\rm 2}$ and keeping $\eta$ to be the same in Eq.~\eqref{eq:tideA}.
As a result, the total tidal correction is
\begin{align}\label{eq:tideBP}
\eta\psi_{\rm BP}  = & \eta \psi_{\rm BP,1} + \eta \psi_{\rm BP, 2} \nonumber \\
= & \lambda_{\rm 1} (q^{-1}\psi^{(-1)}_{\rm BP} + \psi^{(0)}_{\rm BP}  + \sum_{n \ge 1} \eta^n  \psi^{(n)}_{\rm BP})  \nonumber \\
&+ \lambda_{\rm 2} (q \psi^{(-1)}_{\rm BP} + \psi^{(0)}_{\rm BP}  + \sum_{n \ge 1} \eta^n  \psi^{(n)}_{\rm BP})\,.
\end{align}
Strictly speaking, if both compact objects are neutron stars, there is no horizon absorption of the gravitational wave flux. Such effect enters the dynamics at $2.5$ relative PN order for rotating black holes and $4$PN for non-rotating black holes \cite{poisson2004absorption}. The overall contribution to the phase is less than $0.1$  for the point-mass motion terms, which means for the tidal correction it should be even smaller. We shall neglect this effect in the waveform construction. Notice that $\psi^{(0)}_{\rm BP}(\omega)$ becomes the leading order term for star ``2". In fact, it can be evaluated by computing the deformation of a star by an orbiting point mass, and then determining the extra energy change and gravitational wave flux due to the star deformation.
This offers an alternative (and likely easier) way to compute  $\psi^{(0)}_{\rm BP}(\omega)$.

\begin{figure}
	\centering
	\includegraphics[width=0.8\columnwidth, trim=3 7 1 1]{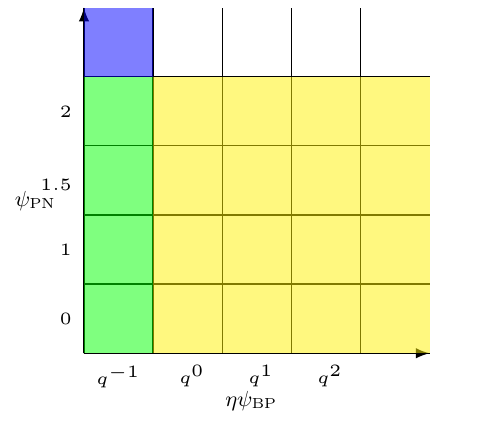}
	\caption{\label{overlap} $\psi_{\rm BP}$ and $\psi_{\rm PN}$ approximately obtained in truncated expansions in the Black Hole Perturbation Theory and Post-Newtonian Formalism. The green and blue parts denote $\psi_{\rm BP}$, the green and yellow parts denote $\psi_{\rm PN}$, and then the overlap of $\psi_{\rm BP}$ and $\psi_{\rm PN}$ is green part.} 
\end{figure}

The tide-induced phase shift can also be expanded in the velocity $v=\sqrt{M/r}$ ($M$ is the total mass and $r$ is the orbital separation) within the PN formalism:
\begin{align}\label{eq:tidePN}
\psi_{\rm PN} =\lambda v^{5} (\psi_{\rm 0PN} +v^2 \psi_{\rm 1PN} + v^3 \psi_{\rm 1.5PN} + v^4 \psi_{\rm 2PN}+...)\,.
\end{align}
Theoretically speaking, after summing over all PN terms in Eq.~\eqref{eq:tidePN} and all mass ratio terms in Eq.~\eqref{eq:tideBP}, $\psi_{\rm BP}$ and $\psi_{\rm PN}$ should agree. In practice, 
$\psi_{\rm BP}$ and $\psi_{\rm PN}$ are approximately obtained in truncated expansions in the Black Hole Perturbation Theory and Post-Newtonian Formalism  respectively, as illustrated in Fig.\ref{overlap}. In order to better capture the tidal effect with these two independent expansions, we propose to construct a hybrid waveform by using
\begin{align}\label{eq:hybrid}
\psi_{\rm hyd} = \psi_{\rm PN} +\psi_{\rm BP} -\psi_{\rm ovp}\,,
\end{align}
where $\psi_{\rm ovp}$ denotes the contribution from the overlap regime of the Post-Newtonian and Black Hole Perturbation methods (the green regime in Fig.\ref{overlap}). As a result, the difference between this hybrid waveform and the ``true" waveform come from the blank space in Fig.~\ref{overlap}. As the expansion orders in Post-Newtonian and Black Hole Perturbation methods increase, the blank space shrinks and we shall obtain a better approximated waveform.  Notice that this construction applies not only to double neutron star and black hole-neutron star binaries discussed here, but black hole binaries as well. It will be interesting to perform the exercise combining the EMRI-inspired waveform with the PN waveform for binary black holes, and compare with 
other resumed waveforms such as the EOB templates.

In this work we truncate the series with only $\psi^{(-1)}_{\rm BP} $ in $\psi_{\rm BP}$ and up to $\psi_{\rm 2PN}$ in $\psi_{\rm PN}$. The accuracy of the resulting hybrid waveform is comparable to the state of the art EOB waveform for the tidal correction, for the numerical waveforms that we have used for comparison. The waveform is naturally expressed in the frequency domain, which allows fast waveform evaluation.  The systematic error is understood as the blank space in the phase diagram as in Fig.~\ref{overlap}. The waveform model is also easily extendible when higher order correction terms in $\psi_{\rm BP}$ and $\psi_{\rm PN}$ are available. We plan to update the hybrid waveform with $\psi^{(0)}_{\rm BP}$ in the future, and possibly with inspiral-to-plunge corrections and higher multipoles if necessary.

The paper is organized as follows. In Section II, we derive the explicit equations of motion of an extended  body with nonzero quadrupole moment moving on a circular and equatorial orbit in the Kerr spacetime. A series of conserved quantities discussed here. In Section III, we review the Teukolsky formalism where the asymptotic behavior of the homogeneous solution, waveforms and fluxes,  and the quadrupole source term are shown. In Section IV, we construct the hybrid waveform and compare it with numerical relativity waveforms, as well as the EOB waveform. We summarize in Section V.
Throughout this paper, we adopt geometrical units, $G=c=1$, where $G$ denotes the gravitational constant and $c$ the speed of light, respectively. The metric signature is $(-,+,+,+)$ 

\section{Conservative  orbital motion}\label{sec:2}
In this section, we consider  a nonspinning  body (with nonzero quadrupolar moment) moving  in the Kerr spacetime, focusing on the case of circular, equatorial orbits. 
Without including the  gravitational radiation reaction, the orbital motion is conservative and easily solvable. We focus on the conservative piece of motion in this section, and leave the discussion on radiative effects to Sec.~\ref{sec:3}.

 The Boyer-Lindquist coordinates $(t,r,\theta,\phi)$ are used in the analysis, in which the Kerr metric takes the following form:
\begin{eqnarray}\label{eq:kerrmetric}
ds^2&=&-\left (1-\frac{2Mr}{\Sigma}\right)\ dt^2-\frac{4aMr {\rm sin}^2\theta}{\Sigma}\ dtd\phi \nonumber\\
 & &+ \frac{\Sigma}{\Delta}\ dr^2  + \Sigma\ d\theta^2 \nonumber\\
 & &+{\rm sin}^2\theta \left(r^2+a^2+\frac{2a^2Mr {\rm sin}^2\theta}{\Sigma} \right)\ d\phi^2,  
\end{eqnarray}
where $M$ is the mass of black hole, $a$ is the spin parameter with $|a|\leq M$, and 
\begin{eqnarray}
\Delta = r^2 - 2 M r + a^2, \quad \Sigma = r^2 + a^2 {\rm cos}^2\theta.
\end{eqnarray}
The Kerr spacetime has two Killing vector fields given by ${\partial_t}$ and ${\partial_\phi}$.

\subsection{Equations of motion}

The motion of a test body with multipolar structure is discussed in detail in \cite{steinhoff2010multipolar}. Following the same formalism,  considering the influence of quadrupole moment-curvature coupling, the equation of motion of a spinning extended body reads 
\begin{align}
\frac{D p_a }{d \tau} &=-\frac{1}{2}R_{abcd}u^{b}S^{cd}-\frac{1}{6} \nabla_a R_{bcde} J^{bcde},  \label{eq:motion1}\\
\frac{DS^{ab}}{d\tau} &= 2p^{[a}u^{b]}+\frac{4}{3}{R^{[a}}_{cde}J^{b]cde}\label{eq:motion2}
\end{align}
where $u^a$ denotes the 4-velocity of the body along its world line (normalized to $u^a u_a=1$), $\tau$ is an affine parameter of the orbit, $R_{abcd}$ denotes the Riemann tensor of a Kerr spacetime, $p^a$ is the momentum, and $J^{abcd}$ is the quadrupole tensor which obeys the following symmetries:
\begin{eqnarray}
J^{abcd} = J^{[ab][cd]} = J^{cdab}, \\
J^{abcd} + J^{bcad} + J^{cabd} = 0.
\end{eqnarray}
 If we only consider the gravito-electric tidal field, neglecting the gravito-magnetic tidal field and quadrupole deformations induced by the spin, the induced quadrupole moment is:
 \begin{eqnarray}\label{eq:quadrumomen}
 	Q^{ab} &=& -\lambda E^{ab},
 \end{eqnarray}
where $\lambda$ is the tidal Love number and $E_{ab}=\frac{1}{{m_1}^2}R_{acbd}p^cp^d$ is the tidal tensor of the spacetime. In addition, the tidal quadrupole deformations $J^{abcd}$ is related to $Q^{ab}$ by
\begin{eqnarray}\label{eq:quadrupole}
J^{abcd} = -\frac{3m_0}{m_1^3}p^{[a} Q^{b][c}p^{d]},
\end{eqnarray}
where
\begin{eqnarray}
	m_{1}^2 &=& -p^{a}p_{a} \,, \nonumber \\
	m_{0} &=& -p_{a}u^{a}\,, \nonumber
\end{eqnarray}
In this paper, we suppose that the extended body has no spin, then the 4-momentum can be obtained from \eqref{eq:motion2}:
\begin{equation}\label{eq:generalized_momentum}
p^a = m_{0} u^a + \frac{4}{3} u_{b} R^{[a}{}_{cde} J^{b]cde}.
\end{equation}
The difference between $m_{0}$ and $m_{1}$ is at higher multipole order than the quadrupole\cite{steinhoff2010multipolar}. As a result, we shall not distinguish  $m_0$ from $m_1$ in this work, as we only consider effects by the quadrupole moment.
The  stress-energy tensor of the test body can be written in the following form:
\begin{eqnarray}\label{eq:t}
T^{ab} = &&\int d \tau \bigg[u^{(a} p^{b)} \frac{\delta_{(4)}}{\sqrt{-g}}+ \frac{1}{3} R_{cde}{}^{(a} J^{b)edc} \frac{\delta_{(4)}}{\sqrt{-g}}\nonumber\\
&&- \frac{2}{3} \nabla_d \nabla_c \big( J^{d(ab)c} \frac{\delta_{(4)}}{\sqrt{-g}}\big) \bigg]\,. \label{em_tensor_singular}
\end{eqnarray}

\subsection{Conserved Quantities}
A test particle moving in the Kerr spacetime has four conserved quantities: energy, angular momentum along the symmetry axis, the Carter constant and its rest mass. As a result, its motion is integrable for generic geodesic orbits. When the internal quadrupole moment is included, we can still construct conserved quantities for extended bodies in the Kerr spacetime based on the Killing vector fields.
According to \cite{ehlers1977dynamics,steinhoff2012influence}, the quantity
\begin{equation}\label{conserved}
Q_{\xi} = p_a \xi^a 
\end{equation}
is conserved if $\xi^a$ is a Killing vector, $\nabla_{(b} \xi_{a)} = 0$. We
then decompose energy and angular momentum as $E =E_0(r) + E_t(r)$ and $J =J_0(r) + J_t(r)$ , where $E_t(r)$ and $J_t(r)$ are proportional to the tidal Love number $\lambda$. As only the first order tidal effects included, we just need to substitute $E_0(r)$ and $J_0(r)$ into Eq.~\eqref{eq:quadrupole} to obtain the momentum $p$ and quadrupole moment $J^{dabc}$ .For the Kerr spacetime, there are two killing fields $\partial_t, \partial_\phi$, which lead to 
\begin{align}\label{E1}
   E=&-p_t \nonumber \\
=&\frac{1}{r}\big(2 M a u^{\phi }-2 M u^t+r u^t\big)-\frac{9M^2\lambda}{m_0^5r^{10}}\big(J_0-E_0a\big)\times\nonumber\\
 &\big(E_0J_0-E_0^2a+m_0^2a\big)\big(2J_0^2-4E_0Ja+2E_0^2a^2+m_0^2r^2\big)\,,\nonumber\\
\end{align}
\begin{align}\label{J1}
	J=&\ p_\phi \nonumber \\
=&\frac{1}{r}\big[2 M a u^t-u^{\phi } \left(2 M a^2+r a^2+r^3\right)\big]\nonumber\\
	&+\frac{9M^2\lambda}{m_0^5r^{10}}\big(J_0-E_0a\big)\big\{2J_{0}^4-6E_0J_0^3a+m_0^2(a^2+r^2)\nonumber\\
	&+J_0^2[6E_0^2a^2+m_0^2(2a^2+3r^2)]\nonumber\\
	&-J_0[2E_0^3a^3+E_0m_0^2a(4a^2+5r^2)]\big\}. 
\end{align}

As both $E, J$ are conserved and the geodesic contributions $E_0, J_0$ are not, one can obtain $E_0, J_0$ at any stage of the orbits as functions of $E,J$ from the above equations. Notice that both $m_{0}$ and $m_1$ are no longer constant with the presence  of quadrupole deformation. In fact, as shown in \cite{steinhoff2012influence}, the following  mass-like quantity $\mu$ as
\begin{equation}\label{masslike}
	\mu=m_{0}+\frac{\lambda}{4}E_{ab}E^{ab}+\mathcal{O}(\lambda^2)
\end{equation} 
 is approximately constant if we neglect the second order tidal effects. It is straightforward to show that Eq.~\eqref{masslike} implies
\begin{align}
 \frac{m_0}{\mu}=&1-\frac{3M^2\lambda}{2m_0^5r^{10}}\big[m_0^4r^4+3m_0^2r^2(J_0-E_0a)^2 \nonumber\\
 &+3(J_0-E_0a)^4\big]+\mathcal{O}(\lambda^2)\,.
\end{align}

\subsection{Orbital description}
In the Kerr spacetime, the motion of a generic test body with internal quadrupolar moment is no longer separable as there are only three conserved quantities: $E$, $J$ and $\mu$. However, for equatorial orbits the inclination angle being a constant: $\theta=\pi/2$, and the motion in $r$ and $\phi$ directions are still separable. In particular, if the orbit is circular, all conserved quantities  can   be expressed as functions of $r$.
With this understanding, we shall explicitly write down the orbital equation of motion up to linear order in $\lambda$ for equatorial orbits.
According to the expressions for $E,J$ in  \eqref{E1}, \eqref{J1} and $m_1^2=m_0^2=-p^{a}p_{a}$, they are 
\begin{align}\label{eq:dttau}
	\bigg(\frac{dt}{d\tau}\bigg)&=\frac{E}{m_0r^2}\bigg[\frac{(r^2+a^2)^2}{\Delta}-a^2\bigg]+\frac{aJ}{m_0r^2}\bigg(1-\frac{r^2+a^2}{\Delta}\bigg)\nonumber\\
	&\quad+F_{0t}(r,a,E,J)  \nonumber\\
	&:=F_{00}(r,a,E,J)+F_{0t}(r,a,E,J),\\
	\bigg(\frac{d\phi}{d\tau}\bigg)&=\frac{ J}{m_0r^2}+\frac{aE}{m_0r^2}\bigg(\frac{r^2+a^2}{\Delta}-1\bigg)-\frac{a^2 J}{m_0r^2\Delta}\nonumber\\
	&\quad+F_{3t}(r,a,E,J)\nonumber\\
	&:=F_{30}(r,a,E,J)+F_{3t}(r,a,E,J),\label{eq:dphidtau}
\end{align}
\begin{widetext}
\begin{align}
	\bigg(\frac{dr}{d\tau}\bigg)^2&=\frac{\big[E(r^2+a^2)-aJ\big]^2}{m_0^4r^4}-\frac{\Delta\big[r^2+(J-aE)^2\big]}{m_0^4r^4}+F_{1t}(r,a,E,J)\nonumber\\
	&:=F_{10}(r,a,E,J)+F_{1t}(r,a,E,J),\label{eq:drdtau}
\end{align}
where
\begin{eqnarray}
	F_{0t}(r,a,E,J)&=&-\frac{9M^2\lambda}{m_0^6 r^{11}\Delta}\bigg(J-Ea\bigg)\bigg[2J^2
	-4EJa+2E^2a^2+m_0^2r^2\bigg]\bigg[2J^2Ma+E^2a(2Ma^2+a^2r+r^3) \nonumber\\
	& &-EJ(4Ma^2+a^2r+r^3)-m_0^2ar(a^2-2Mr+r^2)\bigg], \\
	F_{3t}(r,a,E,J)&=&\frac{9M^2\lambda}{m_0^6 r^{11}\Delta}\bigg(J-Ea\bigg)\bigg[2J^2
	-4EJa+2E^2a^2+m_0^2r^2\bigg]\bigg[-2E^2Ma^2+EJa(4M-r)\nonumber\\
	&&+J^2(-2M+r)+m_0^2r(a^2-2Mr+r^2)\bigg],  \\
	F_{1t}(r,a,E,J)&=&\frac{18 M^2 \lambda}{m_0^7 r^{13}}\bigg(J-Ea\bigg)^2 \bigg[2J^2-4EJa+2E^2 a^2+m_0^2 r^2\bigg]\bigg[-4EJMa+J^2(2M-r)-m_0^2 r\Delta \nonumber\\
	& &+E^2 (2Ma^2 + a^2 r + r^3)\bigg].
\end{eqnarray}
\end{widetext}
The terms $F_{00}, F_{10}, F_{30}$ represent the geodesic motion in the Kerr spacetime, and $F_{0t}, F_{1t}, F_{3t}$ account for the leading-order tidal correction \footnote{ There are no $F_{20}, F_{2t}$ terms here as  the motion in the $\theta$ direction is not present for equatorial orbits.}.
Strictly speaking, the adiabatic tide approximation (Eq.~\eqref{eq:quadrumomen}) breaks down for eccentric orbits as the environmental tidal tensor $E_{ab}$ varies on the orbital timescale. The f-mode excitation and evolution have to be included into the equations of motion \cite{yang2018evolution}. However, as the main purpose of this paper is to generate waveforms for circular orbits, where the adiabatic approximation still holds, we can view Eq.~\eqref{eq:dttau}, Eq.~\eqref{eq:dphidtau} and Eq.~\eqref{eq:drdtau} as effective equations of motions that are introduced as intermediate steps to find the circular orbits.

In the remaining part of the paper, for the sake of convenience, we introduce the following dimensionless variables:
\begin{align}\label{dimenvar}
	r \rightarrow \frac{r}{M},\ J \rightarrow \frac{J}{M\mu},\ E \rightarrow \frac{E}{\mu}
\end{align}
to replace the unnormalized variables.
In this convention, we can rewrite Eq.~\eqref{eq:drdtau} in the form:
\begin{equation}\label{eq:linearmomen}
{\bigg(\frac{1}{r}\frac{dr}{d\tau}\bigg)}^2=\alpha E^2-2\beta\frac{J}{r}E+\gamma\frac{J^2}{r^2}-\delta\frac{m_0^2}{\mu^2},
\end{equation}
where
\begin{eqnarray}
\alpha&=&1+\frac{a^2 (2+r)}{r^3},\ \ \beta = \frac{2a}{r^2}\nonumber\\
\delta&=&1+\frac{a^2-2r}{r^2},\ \ \ \ \gamma=-1+\frac{2}{r}.
\end{eqnarray}
Therefore, we know that Eq.~\eqref{eq:linearmomen} describes a one-dimensional motion within a potential well. For circular orbits, we require the radial velocity to be zero at the equilibrium radius and the radial acceleration to be zero at the same location. Based on these two requirements, we can obtain the conserved $E, J$ as functions of the equilibrium radius $r$: 
\begin{eqnarray}
	E(r)=&&\frac{1-2v^{2}+av^{3}}{\sqrt{1-3v^{2}+2av^{3}}}\nonumber\\
	&&+\lambda \bigg(4r\sqrt{1-3v^2+2av^3}\bigg)^{-1}\bigg[a^2g(r)+2af(r)v\nonumber\\
	&&+\big(-2+r\big)\big(2f(r)+g(r)r\big)\bigg],\label{eq:eqej}\\
	J(r)=&&\frac{1-2v^{2}+av^{3}}{\sqrt{1-3v^{2}+2av^{3}}}\nonumber\\
	&&+\lambda\bigg[4\big(1-3v^2+2av^3\big)\bigg]^{-3/2}\frac{1}{r^{5/2}}\bigg\{2a^4g(r)\nonumber\\
	&&+a^3 v \big[3 g(r) (r-1) r+4 f(r)\big]\nonumber\\
	&&+a^2 \big[g(r) r ((r-1) r-4)+2 f(r) (r-7)\big]\nonumber\\
	&&+3 a \big[g(r) (r-2) (r-1) r+4 f(r)\big]\sqrt{r}\nonumber\\
	&&+(r-3) r^2 \big[g(r) (r-2) r+2 f(r)\big]\bigg\},\label{eq:eqej2}
\end{eqnarray}
where 
\begin{eqnarray}
    v=&&\sqrt{\frac{1}{r}},\nonumber\\
	f(r)=&&\frac{3 \lambda}{r^8 \left(2 a v+r-3\right)^2}  \bigg[-3 a^4+\frac{6 a^3}{v}-a^2 r (3 r+1)\nonumber\\
	&&+\frac{2 a}{v^{5}}-r^2 \left(r^2-3 r+3\right)\bigg],\\
	g(r)=&&-\frac{18 \lambda}{r^9 \left(2 a v+r-3\right)^2}  \bigg[-5 a^4+\frac{12 a^3}{v}-2 a^2 r (2 r+3)\nonumber\\
	&&+\frac{4 a}{v^{5}}-r^2 \left(r^2-2 r+2\right)\bigg]\,.
\end{eqnarray}
In order to compute the gravitational wave fluxes, we also need to evaluate the orbital frequency (only prograde orbits are considered here):
\begin{align}\label{eqomega}
	\Omega_{\phi}&=\frac{d\phi}{dt}=\frac{d\phi/d\tau}{dt/d\tau}=\frac{1}{r^{3/2}+ a}\bigg\{1+\nonumber\\
	&\,\frac{F_{3t}(r,a,E,J)}{F_{30}(r,a,E,J)}-\frac{F_{0t}(r,a,E,J)}{F_{00}(r,a,E,J)}-\nonumber\\
	&\,\frac{(E_tJ_0-E_0J_t)r^2\Delta}{\big[2aE_0+J_0(-2+r)\big]\big[-2aJ_0+E_0r^3+a^2E_0(2+r)\big]}\bigg\}. \nonumber\\ 
\end{align}
If we substitute Eq.~\eqref{eqomega} into Eq.~\eqref{eq:eqej} and Eq.~\eqref{eq:eqej2}, we can obtain $E(\Omega)=E_{0}(\Omega)+E_{t}(\Omega)$ and $J=J_{0}(\Omega)+J_{t}(\Omega)$, where $E_{0}(\Omega)$, $E_{t}(\Omega)$, $J_{0}(\Omega)$ and $J_{t}(\Omega)$ are the geodesics and tidal parts of energy and angular momentum respectively.

We have incorporated these explicit tidal corrections in Eq.~\eqref{eq:eqej} and Eq.~\eqref{eqomega} in an open source Teukolsky code ``Gremlin" within the ``Black Hole Perturbation Toolkit" project\cite{BHPToolkit}, which provides many useful toolboxes for describing the motion and wave emission of EMRIs. This tide-modified Gremlin package allows us to evolve the trajectory of a point particle in a Kerr spacetime, while counting for the tide-induced corrections. In Sec.~\ref{sec:3} we use the same code to compute the gravitational radiation associate with the particle motion.

\subsection{Dynamic tide }

In the low frequency limit, the stars answer to the adiabatic environmental tidal fields by deforming themselves according to Eq.~\eqref{eq:quadrumomen}, with $\lambda$ being a constant. This scenario is often referred as the ``equilibrium tide". In the late part of the inspiral, although the orbital frequency $\Omega$ is still lower than the frequency $\omega_f$ of the $\ell=2$ f-mode, the gradual excitation of the f-mode in the pre-resonance stage is no longer negligible. In fact, as shown in \cite{hinderer2016effects}, effectively we need to replace the constant (dimensionless) Love number
\begin{align}
	\lambda=\frac{2k_2R^5}{3G}
\end{align}
by
\begin{align}
k^{\rm dyn}_{l}=k_{l}\bigg[a_{l}+\frac{b_{l}}{2}\bigg(\frac{Q_{m=l}^{DT}}{Q_{m=l}^{AT}}+\frac{Q_{m=-l}^{DT}}{Q_{m=-l}^{AT}}\bigg)\bigg]
\end{align}
where
\begin{align}
\frac{Q_{m}^{DT}}{Q_{m}^{AT}}&=\frac{\omega_{f}^2}{\omega_{f}^2-(m\Omega)^2}+\frac{\omega_{f}^2}{2(m\Omega)^2\epsilon_{f}\Omega_{f}'(\phi-\phi_{f})}\nonumber\\
&\pm \frac{i\omega_{f}^2}{(m\Omega)^2\sqrt{\epsilon_{f}}}e^{\pm i\Omega_{f}'\epsilon_{f}(\phi-\phi_{f})^2}\int_{-\infty}^{\sqrt{\epsilon_{f}}(\phi-\phi_{f})}e^{\mp i\Omega_{f}'s^2}ds \,.
\end{align}
where the coefficients $a_{2}=1/4$, $b_{2}=3/4$(only $\ell=2$ is considered here), $\Omega^2=M/r^3$ and $\epsilon_{f}$ is the ratio between the orbital timescales and the gravitational radiation reaction timescales, $\Omega'_{f}=3/8$ is a rescaled derivative in frequency, $\phi$ is the orbital phase and $\phi_{f}$ denotes the orbital phase evaluation at $\omega =\omega_f$. These quantities can be written as a function of $r$: 
\begin{align}
    \epsilon_f&=\frac{256\,M^{2/3}\omega_{f}^{5/3}\mu}{5\,|m|^{5/3}}\,, \\
	\phi-\phi_{f}&=\frac{1}{32M^{3/2}\,\mu}\bigg[\bigg(\frac{\sqrt{M}|m|}{\omega}\bigg)^{5/3}-r^{5/2}\bigg]\,,
\end{align}
which  can be found in \cite{hinderer2016effects}. In the above two equations, we do not use the dimensionless variables defined in Eq.~\eqref{dimenvar} in order to express them explicitly.

Note here the star still oscillates at the same frequency of the external tidal force, which is why a frequency-dependent Love number can be introduced here. If the f-mode frequency were within the inspiral frequency range, the post-resonance star also oscillates with a frequency component $\omega =\omega_f$ \cite{yang2019inspiralling}. Such free f-mode oscillations have been observed in numerical simulations of eccentric binary neutron stars \cite{yang2018evolution}.

In Sec.~\ref{sec:4} we show the performance of hybrid waveform models with the dynamic tide effect implemented. The dynamic tide model generally fits better with the numerical waveforms in the late inspiral stage, as demonstrated in \cite{hinderer2016effects}.


\section{Radiation}\label{sec:3}

Neutron stars develop nonzero quadrupole moments because of the gravitational tidal fields from their companions. As a result, the stress-energy of the star is modified by the tidal deformation (Eq.~\eqref{eq:t}). This extra piece of stress energy also generates additional gravitational wave radiation, which in turn affects the orbital evolution. In this section we first review the relevant Teukolsky formalism and then compute the tide-induced gravitational wave radiation.

\subsection{The Teukolsky equation}
The wave emission by an extended body moving in the Kerr spacetime can be described by the Teukolsky equation \cite{teukolsky1973perturbations}, which is separable in the frequency domain.
In particular, consider the variable
\begin{equation}\label{eq:psi4}
\psi_4 = {1\over(r - i a \cos\theta)^4}\int_{-\infty}^{\infty}d\omega
\sum_{lm} R_{lm\omega}(r) _{-2}S^{a\omega}_{lm}(\theta)
e^{im\phi-i\omega t}
\end{equation}
which is a Newmann-Penrose quantity defined by contracting  the Weyl tensor $C_{abcd}$  with tetrad vectors: $\psi_4=-C_{abcd}n^{a}\bar{m}^{b}n^{c}\bar{m}^{d}$. The Kinnersley tetrad components are being used \cite{kinnersley1969type}
\begin{eqnarray}
	n_a=&&\frac{1}{2}(\frac{\Delta}{\Sigma},1,0,-\frac{a\Delta\sin^2\theta}{\Sigma}) \nonumber  \\
	\bar{m}_a=&&\frac{\rho}{\sqrt{2}}(ia\sin\theta,0,\Sigma,-i(r^2+a^2)\sin\theta)
\end{eqnarray}
At any given frequency $\omega$, the wave equation is separable. In particular, the eigen-solution of the angular part of the Teukolsky equation defines 
the spin-weighted spheroidal harmonic $_{-2}S^{a\omega}_{lm}(\theta)$, which is normalized by
\begin{align}
	\int_{0}^{\pi}|_{-2}S^{a\omega}_{lm}(\theta)|^2 \sin\theta d\theta =1.
\end{align}
We have listed relevant properties of the spin-weighted spheroidal harmonics and their derivatives in Appendix A. The radial function $R_{lm\omega}(r)$ obeys the radial Teukolsky equation:
\begin{eqnarray}\label{eq:teukolsky}
\Delta^2 {d\over dr}\left({1\over\Delta}
{dR_{lm\omega}\over dr}\right) - V(r) R_{lm\omega}(r) =
-{\cal T}_{lm\omega}(r).
\end{eqnarray}
where
\begin{eqnarray}\label{eq:potential}
V(r) = -{K^2 + 4i(r-M)K\over\Delta} + 8i\omega r + \lambda_{lm}\;,
\end{eqnarray}
and  $\Delta=r^2-2Mr+a^2$, $K = (r^2+a^2)\omega - m a$,
$\lambda_{lm}\equiv {\cal A}_{lm}
	- 2 a m \omega + a^2 \omega^2 - 2$, where $A_{lm}$ is the eigenvalue of the angular Teukolsky equation.

The radial Teukolsky equation is an ordinary differential equation, which can be solved  by using the Green function method. To achieve this goal, one needs to first identify two independent solutions of the homogeneous Teukolsky equation: $R^H_{lm\omega}$ and $R^\infty_{lm\omega}$, which have the following asymptotic behaviour:
\begin{align}\label{eq:RH}
R^H_{lm\omega} &= B^{\rm trans}_{lm\omega}\Delta^2 e^{-i p_{m} r^*},
\qquad r\to r_+\nonumber\\
R^H_{lm\omega} &= B^{\rm ref}_{lm\omega}r^3 e^{i\omega r^*} +
{B^{\rm inc}_{lm\omega}\over r} e^{-i\omega r^*},\qquad r\to\infty\; 
\end{align}
and
\begin{align}\label{eq:Rinfty}
R^\infty_{lm\omega} &= C^{\rm up}_{lm\omega}e^{i p_{m} r^*} +
C^{\rm ref}_{lm\omega} \Delta^2 e^{-i p_{m} r^*},
\qquad r\to r_+\nonumber\\
R^\infty_{lm\omega} &= C^{\rm trans}_{lm\omega}r^3 e^{i\omega r^*},
\qquad r\to\infty\; .
\end{align}
where $p_{m} = \omega - ma/2Mr_+$ and the tortoise coordinate $r^*$ is:
\begin{equation}
r^*(r) = r + {2M r_+\over r_+ - r_-}\ln{r-r_+\over 2 M}
- {2M r_-\over r_+ - r_-}\ln{r-r_-\over 2 M}, \nonumber\\
\end{equation}
where the outer and inner horizon radii are $r_\pm = M \pm \sqrt{M^2 - a^2}$. Based on  the Green's functions method,  the general solution of the Teukolsky equation with a source can be written in this form:
\begin{equation}\label{eq:teuksolution}
R_{lm\omega}(r) = Z^H_{lm\omega}(r)R^\infty_{lm\omega}(r) +
Z^\infty_{lm\omega}(r)R^H_{lm\omega}(r)\;,
\end{equation}
where
\begin{align}\label{eq:z}
Z^H_{lm\omega}(r) &= \frac{B^{\rm trans}_{lm\omega}}{2i\omega B^{\rm inc}_{lm\omega}
	C^{\rm trans}_{lm\omega}}\int_{r_+}^r dr'
{R^H_{lm\omega}(r'){\cal T}_{lm\omega}(r')\over\Delta(r')^2}\;,\nonumber\\
Z^\infty_{lm\omega}(r) &= {1\over2i\omega B^{\rm inc}_{lm\omega}
	}\int_r^\infty dr'
{R^\infty_{lm\omega}(r'){\cal T}_{lm\omega}(r')\over\Delta(r')^2}\;.
\end{align}

As the neutron star we  consider here moves along circular and equatorial orbits, there is only one frequency in this setup  $\omega=\Omega_\phi$. In particular, the $m$th  harmonic has a frequency of
\begin{align}\label{eq:omegam}
\omega_{m} = m\Omega_\phi.
\end{align}
Then we have 
\begin{align}
	Z^{H,\infty}_{lm\omega}=\sum_{m}\delta(\omega-\omega_{m})Z^{H,\infty}_{lm\omega_{m}}
\end{align}
The energy fluxes going out to infinity  and black hole horizon can be obtained as:
\begin{align}\label{eq:fluxtoih}
\left({dE\over dt}\right)^{r\to\infty}_{GW} &=
\sum_{lm} {{|Z^\infty_{lm\omega_m}|^2}\over4\pi\omega_{m}^2}.\\
\left({dE\over dt}\right)^{r\to r_+}_{GW} &=
\sum_{lm} \alpha_{lm}
{|Z^H_{lm\omega_m}|^2\over4\pi\omega_{m}^2}\;
\end{align}
where
\begin{equation}\label{eq:alpha}
\alpha_{lm} = {256 (2Mr_+)^5 p_{m}(p_{m}^2 +
	4\varepsilon^2)(p_{m}^2 + 16\varepsilon^2)\omega_{m}^3
	\over |C_{lm}|^2}\;,
\end{equation}
with $\varepsilon = \sqrt{M^2 - a^2}/4Mr_+$ and
\begin{align}
|C_{lm}|^2 =& \left[(\lambda_{lm}+2)^2 + 4 a \omega_{m} -
4 a^2 \omega_{m}^2\right][\lambda_{lm}^2 + 36 m a \omega_{m}
\nonumber\\
&- 36 a^2\omega_{m}^2]
 + \left(2\lambda_{lm}+3\right) \left[6 a^2\omega_{m}^2 -
48 m a \omega_{m}\right]\nonumber\\
& + 144 \omega_{m}^2(M^2 - a^2).
\end{align}
As mentioned earlier, the boundary condition for gravitational waves on the star's surface is different from the one for black hole horizon. As a result, the horizon flux should not be accounted for if both objects are neutron stars in the binary system. However, it is a 4PN effect for Schwarzschild black holes  and 2.5PN effect for Kerr black holes, and the associated phase shift is less than $0.1$ even for the point mass motion. Therefore in characterizing the tidal effect and the {\it extra} gravitational wave emission associated with tidal deformation, we shall not consider the issue of the horizon fluxes in our paper.

\subsection{Source term}

In order to obtain the energy flux, we need to evaluate the source term in Eq.~\eqref{eq:z}.
It is explicitly given by \cite{breuer1975gravitational}:
\begin{equation}
{\cal T}_{lm\omega}(r) = 4\int d\Omega\,dt\,{\Sigma\over\rho^4}
\left(B_2' + B_2^{*\prime}\right)_{-2}S^{a\omega}_{lm}(\theta)e^{-im\phi}
e^{i\omega t}\;,
\label{eq:source1}
\end{equation}
where the functions $B_2'$ and $B_2^{*\prime}$ are
\begin{align}
B_2' =& -{\rho^8\bar\rho\over2}L_{-1}\left[\rho^{-4}
L_0\left(\rho^{-2}\bar\rho^{-1} T_{nn}\right)\right] \nonumber\\
&+
{\Delta^2\rho^8\bar\rho\over2\sqrt{2}}L_{-1}\left[\rho^{-4}
\bar\rho^2 J_+\left(\rho^{-2}\bar\rho^{-2}\Delta^{-1}
T_{n\bar m}\right)\right],\nonumber \\
B_2^{*\prime} =& {\Delta^2\rho^8\bar\rho\over2\sqrt{2}} J_+
\left[\rho^{-4}\bar\rho^2 \Delta^{-1} L_{-1}\left(\rho^{-2}
\bar\rho^{-2} T_{n\bar m}\right)\right]\nonumber\\
& -
{\Delta^2\rho^8\bar\rho\over4}J_+\left[\rho^{-4} J_+
\left(\rho^{-2}\bar\rho T_{\bar m\bar m}\right)\right]\;.
\label{eq:B_2^*'}
\end{align}
Here, $\rho = -1/(r - ia\cos\theta)$, $\bar\rho = -1/(r +
ia\cos\theta)$. The differential
operators $J_+$ and $L_{s}$ are
\begin{align}\label{eq:diferentialoperator}
J_+ &= \partial_r + {iK(r)\over\Delta}\;,\nonumber\\
L_s &= \partial_\theta + m\csc\theta - a\omega\sin\theta
+ s\cot\theta\;,\nonumber\\
L_s^{\dagger} &= \partial_\theta - m\csc\theta - a\omega\sin\theta
+ s\cot\theta\;.
\end{align}
The stress-energy tensor for an extended body moving in the Kerr spacetime, as described in Eq.~\eqref{eq:t}, is given by
\begin{align}\label{eq:tab}
	T^{ab}(x) = &\int d\tau \bigg[u^{(a} p^{(b)} \delta_{(4)}+ \frac{1}{3} R_{cde}{}^{(a} J^{b)edc} \delta_{(4)}	\nonumber \\
	&- \frac{2}{3} \nabla_d \nabla_c ( J^{d(ab)c} \delta_{(4)} ) \bigg]. \\
	=&\int d\tau \bigg[u^{(a} p^{(b)} + \frac{1}{3} R_{cde}{}^{(a} J^{b)edc} - J^{daec}\partial_{c}\Gamma^{b}_{de}	\nonumber \\
	&+ \Gamma^{a}_{de}\bigg(\Gamma^{d}_{cf}J^{febc}+\Gamma^{e}_{cf}J^{dfbc}\bigg)\nonumber\\
	&+ \Gamma^{b}_{de}\bigg(\Gamma^{d}_{cf}J^{faec}+\Gamma^{a}_{cf}J^{dfec}\bigg)\bigg]\frac{\delta[x-z(\tau)]}{\sqrt{-g}}\nonumber\\
	&+\partial_{d}\bigg(\Gamma^{d}_{cf}J^{fabc}+\Gamma^{a}_{cf}J^{dfbc}+\Gamma^{b}_{cf}J^{cafd}\bigg)\frac{1}{\sqrt{-g}}\nonumber\\
	&+\partial_{d}\partial_{c}\left\{J^{dabc}\delta[x-z(\tau)]\right\}\frac{1}{\sqrt{-g}}\\
	:=&\int d\tau\,A^{ab}\frac{\delta[x-z(\tau)]}{\sqrt{-g}}+\partial_{d}\left\{B^{bab}\delta[x-z(\tau)]\right\}\frac{1}{\sqrt{-g}}\nonumber\\
	&+\partial_{d}\partial_{c}\left\{J^{dabc}\delta[x-z(\tau)]\right\}\frac{1}{\sqrt{-g}},
\end{align}
where we have converted the covariant derivatives into coordinate partial derivatives with Christoffel symbols, which are more convenient for numerical evaluation. Here the delta function $\delta[x-z(\tau)]$ is defined as
\begin{align}
	\delta [ x-z(\tau)]=\delta [t-t(\tau)]\delta [r-r(\tau)]\delta [\theta-\theta(\tau)]\delta [\phi-\phi(\tau)].
\end{align}
$J^{dabc}$ in Eq.~\eqref{eq:tab} is a tensor, then we have also introduced additional notations for $J^{dabc}$ to account for various pieces of the source terms, as modified by the tidal field
\begin{equation}\label{eq:Ann1}
A_{\{nn\}}:=A^{ab}n_{a}n_{b},
\end{equation}
\begin{equation}\label{eq:BJ1}
B^{\{\alpha\}}_{\{nn\}} := B^{\alpha ab}n_{a}n_{b},\ \ \ \ J^{\{\alpha\beta\}}_{\{nn\}}:=J^{\alpha ab\beta}n_{a}n_{b}.
\end{equation}
where $\alpha,\beta=t,r,\theta,\phi$, and
\begin{align}
B^{\{r\}}_{\{dr nn\}} &:= B^{r ab}\partial_{r}(n_{a}n_{b})\,,  \ \ B^{\{\theta\}}_{\{d\theta nn\}} := B^{\theta ab}\partial_{\theta}(n_{a}n_{b}),\nonumber\\ 
J^{\{tr\}}_{\{dr nn\}}&:=J^{t abr}\partial_{r}(n_{a}n_{b})\,, \ \ J^{\{t\theta\}}_{\{d\theta nn\}}:=J^{t ab\theta}\partial_{\theta}(n_{a}n_{b}),\nonumber\\
J^{\{r\phi\}}_{\{dr nn\}}&:=J^{r ab\phi}\partial_{r}(n_{a}n_{b})\,,\ \ J^{\{\theta\phi\}}_{\{d\theta nn\}}:=J^{\theta ab\phi}\partial_{\theta}(n_{a}n_{b}),\nonumber\\
J^{\{rr\}}_{\{ddr nn\}}&:=J^{r abr}\partial_{r}\partial_{r}(n_{a}n_{b})\,,\ J^{\{\theta\theta\}}_{\{dd\theta nn\}}:=J^{\theta ab\theta}\partial_{\theta}\partial_{\theta}(n_{a}n_{b}).\nonumber\\
J^{\{r\theta\}}_{\{dr nn\}}&:=J^{r ab\theta}\partial_{r}(n_{a}n_{b})\,, \ \
J^{\{r\theta\}}_{\{d\theta nn\}}:=J^{r ab\theta}\partial_{\theta}(n_{a}n_{b})\,\nonumber\\
J^{\{r\theta\}}_{\{drd\theta nn\}}&:=J^{r\theta ab\theta}\partial_{r}\partial_{\theta}(n_{a}n_{b}).
\end{align}
Apart from $n_{a}n_{a}$, we can also define components for $J^{d ab c}$ by contracting $J^{d ab c}$ with $n_{(a}\bar{m}_{b)}$ and $\bar{m}_{a}\bar{m}_{b}$ following similar convention as the above equations. The explicit forms of these components are given in the Appendix B.

\subsection{Sample evolution}\label{sec:3c}

\begin{figure}
	\includegraphics[width=0.9\columnwidth,trim=0 7 0 0]{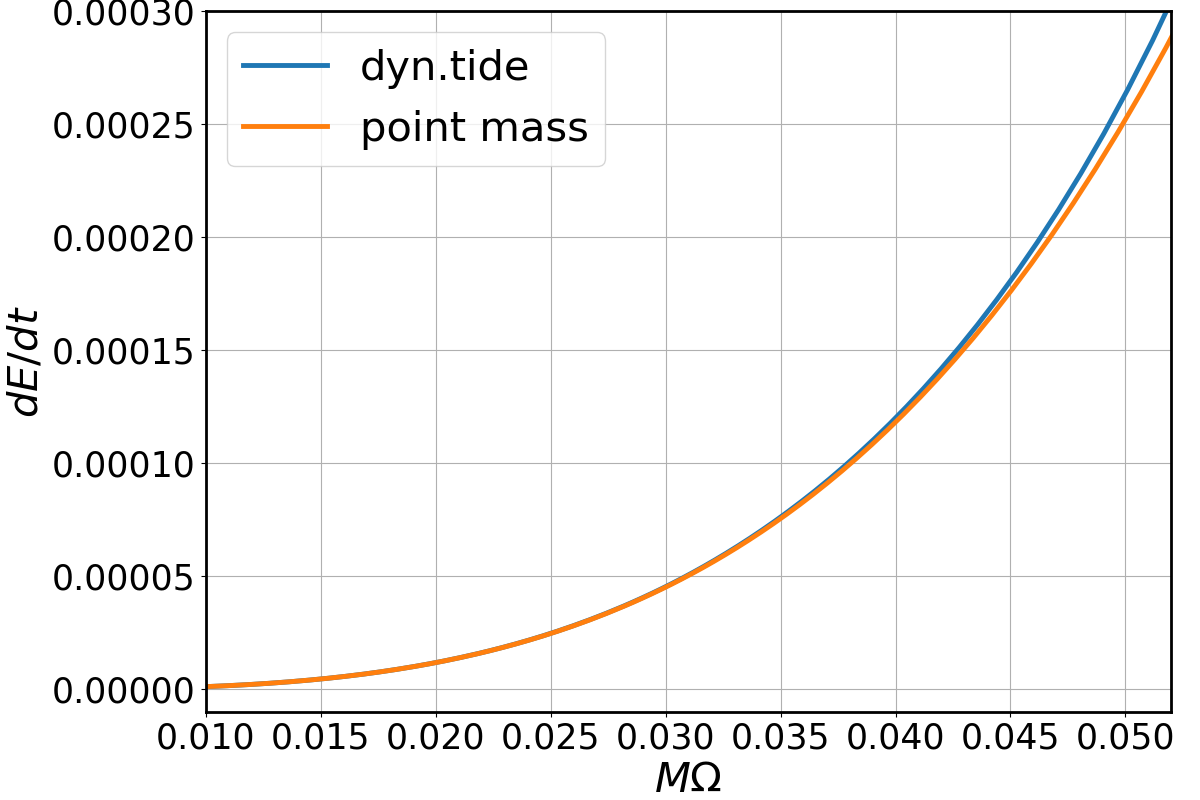}
	\caption{\label{GWflux}  The energy flux computed for an equal-mass, black hole-neutron star binary with $k_2(\Omega=0)=0.07346,\ m_0=1.4M_{\odot},\ M\omega_f = 0.1349(1+q)/2 $ and $\Gamma=2$ polytropic equation of state, and a similar binary black hole system with the same mass and starting from the same initial location. The flux $dE/dt$ is normalized by $\eta^2$.}
\end{figure}

We incorporated the additional tide-related source terms into the Gremlin code, and evaluate the gravitational wave energy flux as a function of the orbital frequency. Formally we can write the total power as
\begin{align}\label{eq:dedt}
P = P_{\rm pm}+P_{\rm tide} =\eta^2 \left({dE\over dt}\right)^{\rm pm}+ \eta \frac{\lambda}{m^5_0} \left({dE\over dt}\right)^{\rm tide}\,.
\end{align}
The $\eta^2$ factor within the point mass term is related to the fact that metric perturbation generated by the point mass is proportional to the mass ratio, so that the flux is proportional to $\eta^2$.
The tidal correction of the gravitational wave flux is generated by the beating of the wave generated by the point mass with the additional wave generated by the quadrupole deformation of the star.
Both $\dot{E}^{\rm pm}$ and $\dot{E}^{\rm tide}$ can be computed given the initial conditions of the system. The values can be used in other systems with different $\eta$ and $\lambda$.

In Fig.~\ref{GWflux}, we plot the total power versus the point mass power for a non-spinning,  equal-mass binary neutron star system. The same type of system is also used in Sec.~\ref{sec:4} for waveform comparison. The additional energy flux contributed by the tidal deformation (Eq.~\eqref{eq:dedt}) becomes more important at higher frequencies.
Although the fluxes are computed within the extreme-mass-ratio limit, the results are applied in the comparable mass ration limit for the waveform construction.

\section{Waveform Construction}\label{sec:4}

With the preparation in Sec.~\ref{sec:2} and Sec.~\ref{sec:3} on the conservative and dissipative pieces of the tidal effects, we are ready to present  the tidal correction to the gravitational waveform.
We shall focus on the gravitational wave phase as it is the most sensitively measured quantity within a parameter estimation process.

Assuming adiabatic circular orbit evolution, the motion at any instantaneous moment can be approximately viewed as a circular orbit with frequency $\Omega$. The gravitational wave phase, as a function of the orbital frequency, follows
\begin{align}\label{eq:orbitalphase}
	\frac{d^2 \psi}{d\Omega^2}&=2\frac{dE/d\Omega}{P}\,.
\end{align}
As we are interested in the tidal correction, we shall write the total phase $\psi$ as $\psi^{\rm pm}+\psi^{\rm tide}$, the total energy as $E= E^{\rm pm} + E^{\rm tide}$, and expand Eq.~\eqref{eq:orbitalphase} so that only linear order terms in $\lambda$ are kept:
\begin{align}\label{eq:dphitide}
\frac{d^2 \psi^{\rm tide}}{d \Omega^2} = 2 \left ( \frac{dE^{\rm tide}/d\Omega}{P} -P^{\rm tide} \frac{dE/d\Omega}{P^2} \right )\,,
\end{align}
where we plug in $E^{\rm tide}$ and $P^{\rm tide}$ evaluated in Sec.~\ref{sec:2} and Sec.~\ref{sec:3}.
In the Post-Newtonian theory, $E^{\rm tide}$ and $P^{\rm tide}$ can be computed to various PN orders, which lead to the PN tide waveform at different orders \cite{vines2011post}.
Notice that the gravitational wave phase increases twice as fast as the orbital phase, because we focus on the dominant piece of the waveform with $\ell=2, m=2$.

\subsection{Hybrid waveform}

The black hole perturbation calculation discussed in Sec.~\ref{sec:2} and Sec.~\ref{sec:3} gives rise to an  EMRI-inspired waveform, which is fully capable of  describing the gravitational wave emission in the highly relativistic regime. On the other hand, the PN tide waveform, although being less accurate in the strong-gravity regime, does not require an expansion in the mass ratio. In order to combine the merits of these two different approaches, we have proposed a hybrid version of the waveform, as explained in Eq.~\eqref{eq:hybrid} and depicted in Fig.~\ref{overlap}. By definition, this hybrid waveform is accurate if the mass ratio is small {\it or} if the binary separation is large. Similar to the spirit of the EOB construction, we anticipate that by ensuring  matching at small mass ratio and weak gravity limit, the hybrid method still provides reasonably accurate description for comparable  mass-ratio systems in the strong gravity regime. This point has to be checked with numerical relativity waveforms, as discussed in Sec.~\ref{sec:4b}.

In constructing the hybrid waveform one needs to subtract the waveform contribution in the overlap regime, as explained in Fig.~\ref{overlap}. In fact, it also serves as a sanity check that the PN waveform taking a mass ratio expansion should agree with the EMRI-inspired waveform taking a PN expansion. In light of Eq.~\eqref{eq:dphitide}, it suffices to show that $E^{\rm tide}$ and $P^{\rm tide}$ obtained in the PN theory have the same small mass limit as their counterparts found in Sec.~\ref{sec:2} and Sec.~\ref{sec:3}, expanded in various PN orders. Such a consistency check is explicitly performed in Appendix.~\ref{ap:c}.

\subsection{Numerical comparison}\label{sec:4b}

In order to evaluate the performance of the black hole perturbation and hybrid methods in constructing waveforms, we adopt an equal mass, binary neutron star waveform from the SXS waveform catalog \cite{boyle2019sxs}.
For this particular waveform,  the neutron stars have a polytropic equation of state $P=K\rho^{\Gamma}$, with $\Gamma=2$, $K=101.45$. The neutron star mass is $m=1.4M_{\odot}$ and the radius is $R=14.4$km. The phase error is approximately $\sim 1$ rad at the peak of the strain \cite{hinderer2016effects}.

\begin{figure}
	\includegraphics[width=0.9\columnwidth,trim=0 7 0 0]{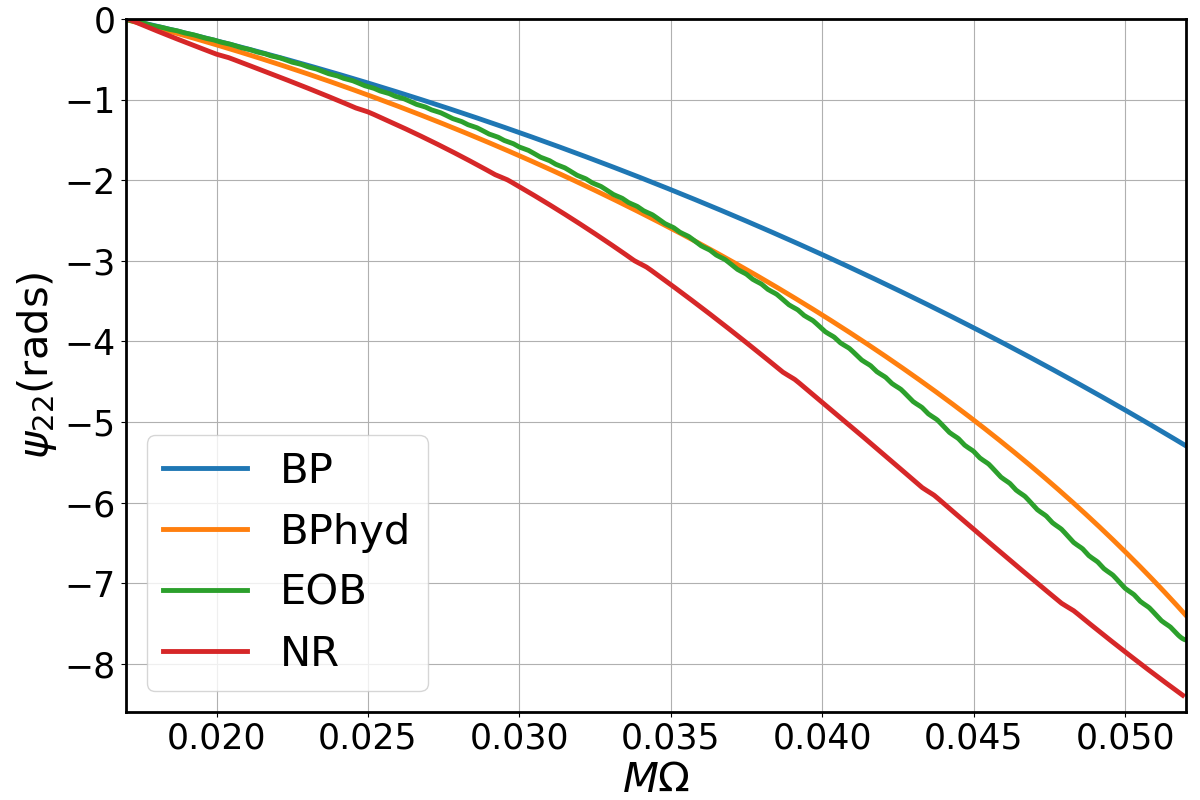}
	\caption{\label{deltaphase}  Tidal phases of gravitational perturbation theory, the hybrid method, the EOB framework and numerical relativity simulation for an equal-mass, black hole-binary neutron system. The property of neutron star is the same as Fig.~\ref{GWflux}. } 
\end{figure}

\begin{figure}
	\includegraphics[width=0.9\columnwidth,trim=0 7 0 0]{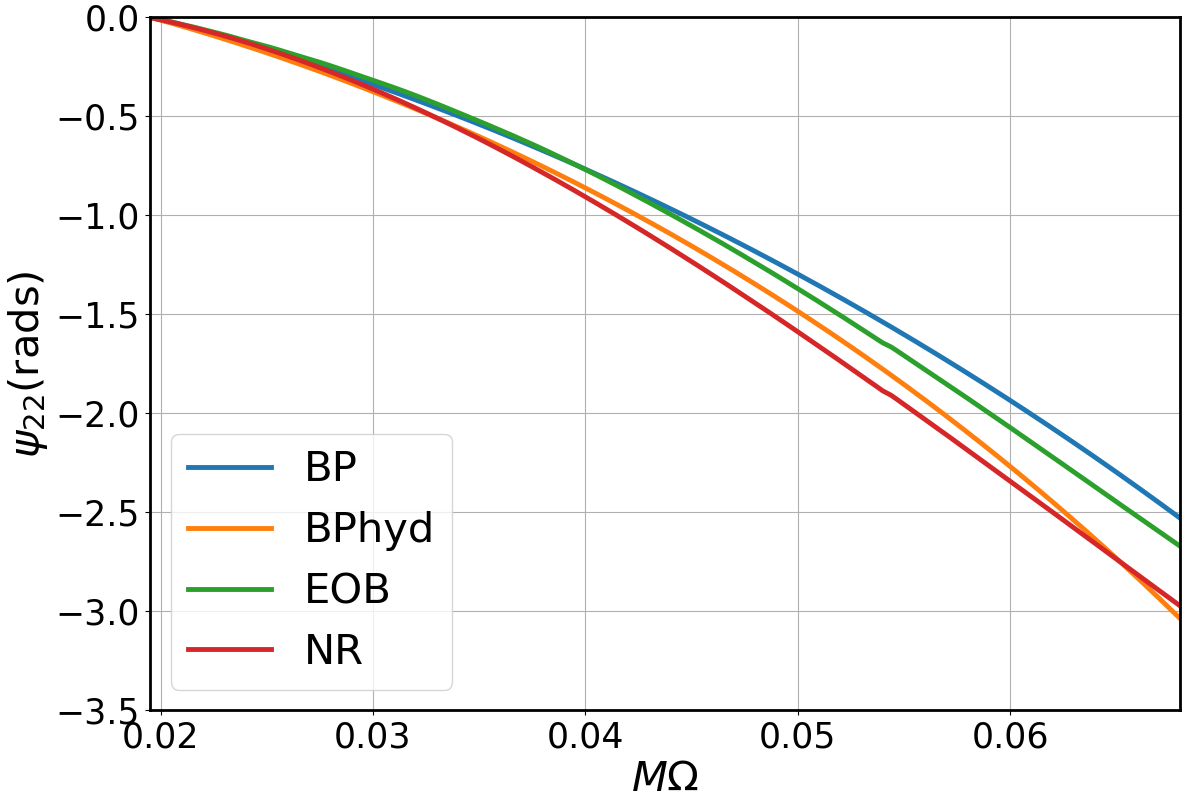}
	\caption{\label{deltaphase2}  Tidal phases of gravitational perturbation theory, the hybrid method, the EOB framework and numerical relativity simulation for black hole-binary neutron system with mass ratio 2:1. The property of neutron star is the same as Fig.~\ref{GWflux}.} 
\end{figure}

For comparison purpose, we also compute the EOB  prediction of the tidal phase correction, with dynamic tide included, in addition to the black hole perturbation result. As shown in Fig.~\ref{deltaphase}, the hybrid waveform that integrates both the black hole perturbation and 2PN  methods, performs significantly better than the black hole perturbation result alone. This hybrid waveform also has comparable performance as the EOB-dynamic tide waveform. In Fig.~\ref{deltaphase2}, we consider a black hole-neutron star system with mass raio 2:1 and the property of he neutron star is the same as Fig.~\ref{GWflux} and Fig.~\ref{deltaphase}. We observe slightly better agreement with the numerical wavefrom for the hybrid waveform is  in this case, but the difference is within the phase uncertainty of the numerical waveform. Apart from these two scenarios, more detailed and systematic comparison and characterization are needed to address the phase error of the hybrid waveform. 

This hybrid waveform is naturally expressed in the frequency domain, which is convenient for fast waveform evaluation. To further improve the waveform accuracy to meet the requirements of third-generation gravitational wave detectors, high-order corrections ($q^0$ and $\eta^1$) in the black hole perturbation method should be evaluated to reduce the empty space in Fig.~\ref{overlap}. As numerical waveforms are required for validation and calibration purposes, we also likely require future numerical waveforms with $\mathcal{O}(0.1)$ phase error, i.e., a factor of ten improvement from current waveforms.

\begin{figure}
	\includegraphics[width=0.88\columnwidth,trim=0 7 0 0]{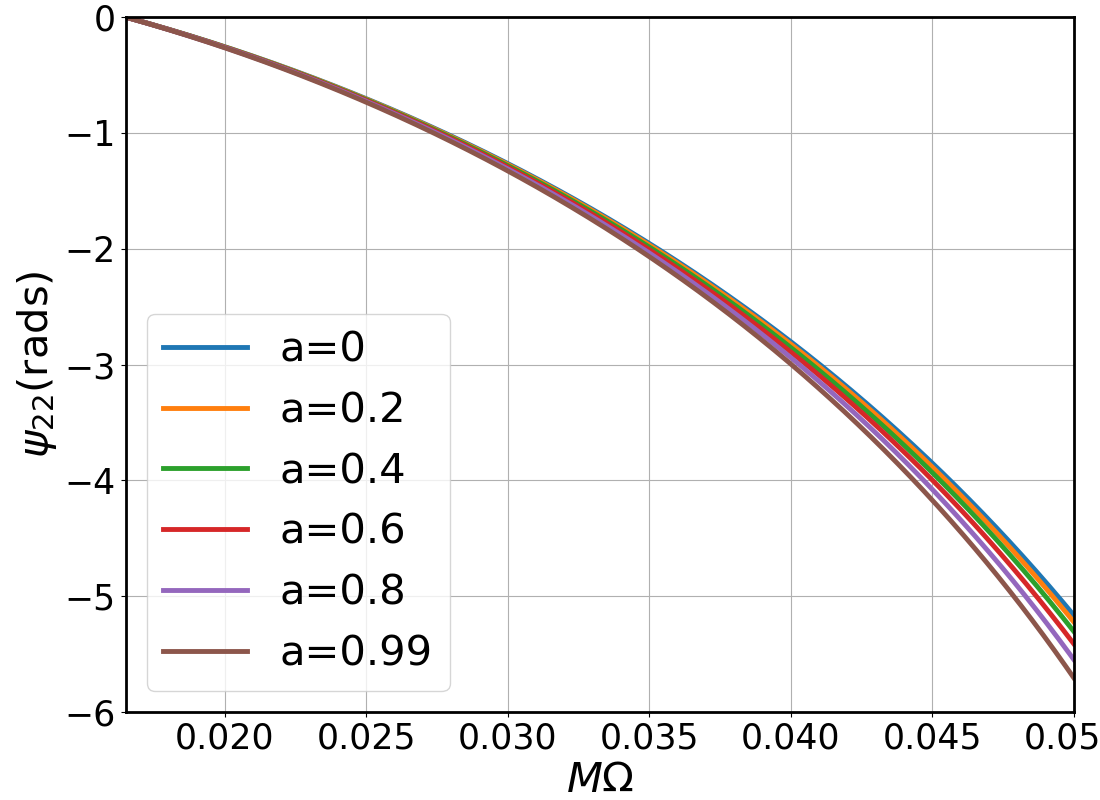}
	\caption{\label{spin}  Tidal phases in the black hole perturbation waveform with spin ranging from $a=0$ to $a=0.99$ for six equal-mass, black hole-neutron star systems. The property of neutron star is the same as Fig.~\ref{GWflux}.} 
\end{figure}

Interestingly, the black hole perturbation approach also offers straightforward evaluation of the spin-dependence of the tidal terms, which are absent in the current PN or EOB waveforms. According to Fig.~\ref{spin},  the influence of the spin parameter of the black hole on the tide-induced phase shift is less than 10\% in the entire inspiral range. Such additional phase shift may be less important for binary neutron stars, as they are generally believed to be slowly spinning according to the observation of galactic pulsar binaries \cite{andrews2019double}. Nevertheless they should be relevant for black hole-neutron star binaries if we want to control the waveform phase error to be below 0.1, especially for the ones with a low-mass black hole \cite{yang2018can}. 


\section{conclusion}

A recent program in connecting scattering amplitude calculations to two-body problems in General Relativity has triggered an evolution in Post-Newtonian and Post-Minkowski (PM) Theory \cite{damour2020classical,bini2020sixth,bini2020binary}. 
Higher order PN and PM corrections to the equations of motion have been discovered with this new approach.
On the other hand, the development of second-order (in mass ratio) gravitational self force is being carried out and implemented in circular orbits in Schwarzschild \cite{pound2020second}. It is expected to correct phase error on the $\mathcal{O}(1)$ order, which is on the comparable level of the environmental effects \cite{yang2017general,bonga2019tidal,yunes2011imprint,barausse2014can}.
The hybrid approach proposed here naturally integrates these two independent expansions to achieve a better description of binary motions in the comparable mass ratio, strong gravity and high velocity regime.
In this work we have incorporated the PN expansion of the tidal correction up to 2PN order and the leading-order term in the mass-ratio expansion, which gives rise to a hybrid waveform with comparable accuracy to the state of the art EOB waveform.

Moving forward, it should be straightforward to include the $\psi_{\rm 2.5PN}$ term \cite{agathos2015constraining} and $\psi^{(0)}_{\rm 1BP}$ corrections.
In particular, as $\psi^{(0)}_{\rm 1BP}$ is the leading-order tidal term for the more massive object, it is easier to consider the problem of a point mass orbiting around a star to evaluate  $\psi^{(0)}_{\rm 1BP}$.
In Fig.~\ref{spin}, we observe that the discrepancy between the black hole perturbation waveform and the numerical waveform monotonically increases as the binary evolves.
The inclusion of $\psi_{\rm 2.5PN}$ term and $\psi^{(0)}_{\rm 1BP}$ may help alleviate the disagreement. 
In the future, it is feasible to also work out the $\psi^{(1)}_{\rm 1BP}$ and beyond-2PN corrections to achieve better accuracy.

In  \cite{van2020intermediate}, it is argued that for the comparison  between the numerical relativity binary black hole waveform and the EMRI-inspired waveform, the discrepancy at large orbital frequency might come from the breakdown of the adiabatic approximation, so that the inspiral-to-plunge transition has to be taken into account.
However, in the comparison performed here for the tidal effects, the discrepancy never displays a sudden rise near the merger. Therefore we do not expect the inspiral-to-plunge transition to be the main reason of disagreement found here. Nevertheless, we may still include the transition in our future investigation to see how it affects the waveform construction.

On the other hand, higher-order in mass ratio terms may be obtained by calibration with a set of numerical waveforms \cite{van2020intermediate}. Let us consider Eq.~\eqref{eq:tideBP} as an example. If both $\psi^{(-1)}_{\rm 1BP}$  and $\psi^{(0)}_{\rm 1BP}$ are known through black hole perturbation calculations, we may truncate the summation up to $n \ge 1$ and fit $\psi^{(1)}_{\rm 1BP},...,\psi^{(n)}_{\rm 1BP}$ by comparing to a series numerical waveforms with different mass ratios.  The obtained fitting formula and the associated waveform can be tested with another independent set of numerical waveforms. The accuracy of this method relies crucially on the accuracy of the calibration waveforms.
We plan to perform this analysis using more binary neutron star and black hole-neutron star waveforms.

As Advanced LIGO continues to improve its sensitivity and especially with the third-generation gravitational-wave detectors \cite{hild2011sensitivity,reitze2019cosmic}, we should expect to observe a set of  high signal-to-noise-ratio (SNR) events, which will allow many important applications of precise gravitational wave astronomy. The gain in SNR also poses strict requirements on the modeling error of the waveforms, so that the waveform systematic error is smaller than the statistical error of these events. It has been shown that for third generation detectors the mismatch error for numerical relativity waveforms has to improve by one order of magnitude. For semi-analytical waveforms an improvement of three orders of magnitude is necessary \cite{purrer2020gravitational}. Significant new developments are required to bridge such a large gap, and hopefully the hybrid method proposed here will provide one avenue for exploration.

\acknowledgements
We thank Tanja Hinderer for helping with  the EOB waveform and Adam Pound for interesting discussions. We thank B\'{e}atrice Bonga for reading over the manuscript and providing many helpful comments. This work makes use of the Black Hole Perturbation Toolkit. This work was supported by the Strategic Priority Research Program of the Chinese Academy of Sciences under Grants No.XDA1502070401. H. Y. is supported by the Natural Sciences and
Engineering Research Council of Canada and in part by
Perimeter Institute for Theoretical Physics. Research at
Perimeter Institute is supported in part by the Government of
Canada through the Department of Innovation, Science and
Economic Development Canada and by the Province of Ontario through the Ministry of Colleges and Universities.


\appendix

\section{Spheroidal harmonics}
Even though some derivatives of the spin-weighted spheroidal harmonics can be found in \cite{hughes2000evolution}, we need some other derivatives which we state them as follows:
\begin{eqnarray}
\partial_{\theta}{_{-1}Y}_{km}(\theta)&&=(m\csc\theta-\cot\theta){_{-1}Y}_{km}(\theta)\nonumber\\
&&-[k(k+2)]^{1/2}{_{0}Y}_{km}(\theta),
\end{eqnarray}
\begin{eqnarray}
\partial_{\theta}{_{0}Y}_{km}(\theta)&&=m\csc\theta{_{0}Y}_{km}(\theta)\nonumber\\
&&-[k(k+2)]^{1/2}{_{1}Y}_{km}(\theta),
\end{eqnarray}
\begin{eqnarray}
\partial_{\theta}{_{1}Y}_{km}(\theta)&&=m\csc\theta{_{1}Y}_{km}(\theta)\nonumber\\
&&-[(k-1)(k+2)]^{1/2}{_{2}Y}_{km}(\theta),
\end{eqnarray}
\begin{eqnarray}
S=\sum_{k=l_{min}}^{\infty}b_{k}\partial_{\theta}{_{-2}Y}_{km},
\end{eqnarray}
\begin{eqnarray}
\partial_{\theta}S =&& (m\cot\theta\csc\theta+2\csc^2\theta)S\nonumber\\
&&-\sum_{k=l_{min}}^{\infty}b_{k}[(k-1)(k+2)]^{1/2}{_{-1}Y}_{km}(\theta),
\end{eqnarray}
\begin{align}
\partial_{\theta}\partial_{\theta}S =& (-m\cot\theta\csc\theta+2\csc^2\theta)S\nonumber\\
&+(m\cot\theta\csc\theta+2\csc^2\theta)\partial_{\theta}S\nonumber\\
&-\sum_{k=l_{min}}^{\infty}b_{k}[(k-1)(k+2)]^{1/2}\partial_{\theta}{_{-1}Y}_{km}(\theta),
\end{align}
\begin{align}
\partial_{\theta}L_{1}^{\dagger}L_{2}^{\dagger}S =& \sum_{k=l_{min}}^{\infty}b_{k}[(k-1)k(k+1)(k+2)]^{1/2}\partial_{\theta}{_{0}Y}_{km}(\theta)\nonumber\\
&+2a\omega\cos\theta L_{2}^{\dagger}S+2a\omega\sin\theta\partial_{\theta}L_{2}^{\dagger}S\nonumber\\
&-2(a\omega\sin\theta)a\omega\cos\theta S-(a\omega\sin\theta)^2\partial_{\theta}S,
\end{align}
\begin{align}
L_{2}^{\dagger}S =& a\omega\sin\theta S-\sum_{k=l_{min}}^{\infty}b_{k}[(k-1)(k+2)]^{1/2}{_{-1}Y}_{km}(\theta)\\
\partial_{\theta}L_{2}^{\dagger}S =& a\omega\cos\theta S+ a\omega\sin\theta\partial_{\theta}S\nonumber\\
&-\sum_{k=l_{min}}^{\infty}b_{k}[(k-1)(k+2)]^{1/2}\partial_{\theta}{_{-1}Y}_{km}(\theta),\\
L_{1}^{\dagger}L_{2}^{\dagger}S =& \sum_{k=l_{min}}^{\infty}b_{k}[(k-1)k(k+1)(k+2)]^{1/2}{_{0}Y}_{km}(\theta)\nonumber\\
&+2a\omega\sin\theta L_{2}^{\dagger}S-(a\omega\sin\theta)^2 S
\end{align}
\begin{align}
\partial_{\theta}\partial_{\theta}L_{2}^{\dagger}S =&(-a\omega\sin\theta+2a\omega\sin\theta)S\nonumber\\
&+(a\omega\cos\theta+a\omega m-2a\omega\cos\theta)\partial_{\theta} S\nonumber\\
&-(a\omega\cos\theta-m\cot\theta\csc\theta+\csc^2\theta)\times\nonumber\\
&\sum_{k=l_{min}}^{\infty}b_{k}[(k-1)(k+2)]^{1/2}{_{-1}Y}_{km}(\theta)\nonumber\\
&-(a\omega\sin\theta+m\csc\theta-\cot\theta)\times\nonumber\\
&\sum_{k=l_{min}}^{\infty}b_{k}[(k-1)(k+2)]^{1/2}{_{-1}Y}_{km}\nonumber\\
+&\sum_{k=l_{min}}^{\infty}b_{k}[(k-1)k(k+1)(k+2)]^{1/2}{_{0}Y}_{km},
\end{align}
\begin{align}
\partial_{\theta}\partial_{\theta}L_{1}^{\dagger}L_{2}^{\dagger}S =& (-m\cot\theta\csc\theta)\times\nonumber\\
&\sum_{k=l_{min}}^{\infty}b_{k}[(k-1)k(k+1)(k+2)]^{1/2}{_{0}Y}_{km}(\theta)\nonumber\\ &+m\csc\theta\sum_{k=l_{min}}^{\infty}b_{k}[(k-1)(k+2)]^{1/2}\partial_{\theta}{_{0}Y}_{km}(\theta)\nonumber\\
&-\sum_{k=l_{min}}^{\infty}b_{k}[(k-1)k^2(k+1)^2(k+2)]^{1/2}\partial_{\theta}{_{1}Y}_{km}(\theta)\nonumber\\
&-2a\omega\sin\theta L_{2}^{\dagger}S+2a\omega\cos\theta\partial_{\theta}L_{2}^{\dagger}S\nonumber\\
&+2a\omega\cos\theta\partial_{\theta}L_{2}^{\dagger}S+2a\omega\sin\theta\partial_{\theta}\partial_{\theta}L_{2}^{\dagger}S\nonumber\\
&-2a^2\omega^2(\cos^2\theta-\sin^2\theta)S\nonumber\\
&-2a^2\omega^2\sin\theta\cos\theta\partial_{\theta}S-2a^2\omega^2\sin\theta\cos\theta\partial_{\theta}S\nonumber\\
&-(a\omega\sin\theta)\partial_{\theta}\partial_{\theta} S.
\end{align}

\begin{widetext}
\section{Source terms}\label{sourcecomponents}
Because we consider the first order tidal effects, the we can substitute $E_0$ and $J_0$ into the tensors $B^{dab}$ and $J^{dabc}$. The concrete components are
\begin{align}
A_{\{nn\}}=&\frac{\left[Ja-E(r^2+a^2)\right]^2}{4mr^4}+\frac{\lambda(a^2+(-2+r)r)^2}{4r^{11}(-3+2av+r)^3}\bigg[-18a^4+\frac{a^3}{v}\nonumber\\
&-2a^2r(31+3r)+\frac{4a(3+4r)}{v^3}+r^2(9-12r+r^2)\bigg]
\end{align}
\begin{align}
A_{\{n\bar{m}\}}=&-\frac{i (J-E a) \left[J a-E \left(r^2+a^2\right)\right]}{2 \sqrt{2}m r^3}+\frac{i \lambda}{2 \sqrt{2} \left(2 v a+r-3\right)^3} \left(\frac{1}{r}\right)^{19/2} \bigg[18 v a^7\nonumber\\
&+\frac{(33 r+17) a^5}{v}+r (133-97 r) a^4+\frac{\left(17 r^2+48 r-139\right) a^3}{v^{3}}+r^2 \left(-39 r^2+84 r-17\right) a^2\nonumber\\
&+\frac{\left(2 r^3+21 r^2-83 r+66\right) a}{v^{5}}+r^3 \left(-2 r^3+r^2+15 r-18\right)-60 a^6\bigg],
\end{align}
\begin{align}
A_{\{\bar{m}\bar{m}\}}=&\frac{E \left[(r+2) a^2+r^3\right]-2 J a}{mr \left[(r-2) r+a^2\right]}-\frac{\lambda}{r^9 \left(2 v a+r-3\right)^3}\bigg[  \frac{30 a^5}{v}-r (12 r+7) a^4+\frac{4 (11 r-20) a^3}{v^{3}}\nonumber\\
&-r^2 \left(r^2+46 r-104\right) a^2+\frac{2 \left(4 r^2-2 r-15\right) a}{v^{5}}+r^3 \left(r^3-10 r^2+21 r-9\right)-9 a^6 \bigg],
\end{align}
\begin{align}
B^{\{t\}}_{\{nn\}}=&\frac{3 \lambda  \big[(r-2) r+a^2\big]} {8 r^9 \left(2 v a+r-3\right)^2}\bigg[6 v a^5+6 r a^4+\frac{(3 r-25) a^3}{v}\nonumber\\
&+2 r \left(3 r^2-5 r+5\right) a^2+\frac{3 \left(r^2-5 r+4\right) a}{v^{3}}+2 r^3 \left(r^2-5 r+6\right)\bigg],
\end{align}
\begin{align}
B^{\{t\}}_{\{n\bar{m}\}}=&-\frac{3 i \lambda  }{8 \sqrt{2} r^8 \left(2 v a+r-3\right)^2}\bigg[6 v a^6+6 (r-1) a^5-\frac{(3 r+19) a^4}{v}\nonumber\\
&+2 r \left(3 r^2-2 r+13\right) a^3-\frac{2 (4 r+5) a^2}{v^{3}}+2 r^2 \left(r^3-2 r^2+3\right) a+\frac{r^2-5 r+6}{v^{7}}\bigg],
\end{align}
\begin{align}
B^{\{t\}}_{\{\bar{m}\bar{m}\}}=0,
\end{align}
\begin{align}
B^{\{r\}}_{\{nn\}}=&\frac{3 \lambda  \big[(r-2) r+a^2\big]^2 }{8 r^{10} \left(2 v a+r-3\right)^2}\bigg[\frac{6 a^3}{v}-r (3 r+1) a^2+\frac{2 a}{v^{5}}-r^2 \left(r^2-3 r+3\right)-3 a^4\bigg],
\end{align}
\begin{align}
B^{\{r\}}_{\{n\bar{m}\}}=&\frac{3 i \lambda  v^{17} \big[(r-2) r+a^2\big]}{8 \sqrt{2} \left(2 v a+r-3\right)^2}\bigg[6 v a^5+\frac{2 (3 r+4) a^3}{v}\nonumber\\
&+r (15-13 r) a^2+\frac{2 \left(r^2-3\right) a}{v^{3}}-3 r^2 \left(r^2-3 r+2\right)-18 a^4\bigg],
\end{align}
\begin{align}
B^{\{r\}}_{\{\bar{m}\bar{m}\}}=&-\frac{3 \lambda  }{4 r^8 \left(2 v a+r-3\right)^2}\bigg[\frac{12 a^5}{v}-r (3 r+10) a^4+\frac{(11 r-15) a^3}{v^{3}}\nonumber\\
&-r^2 \left(r^2+9 r-25\right) a^2+\frac{\left(3 r^2-5 r-6\right) a}{v^{5}}+r^3 \left(-2 r^2+6 r-3\right)-3 a^6\bigg],
\end{align}
\begin{align}
B^{\{r\}}_{\{dr nn\}}=&\frac{3 \lambda  \left(r-a^2\right) \big[(r-2) r+a^2\big]}{2 r^{11} \left(2 v a+r-3\right)^2}\bigg[\frac{6 a^3}{v}-r (3 r+1) a^2\nonumber\\
&+\frac{2 a}{v^{5}}-r^2 \left(r^2-3 r+3\right)-3 a^4\bigg],
\end{align}
\begin{align}
B^{\{r\}}_{\{dr n\bar{m}\}}=&-\frac{3 i \lambda  }{8 \sqrt{2} r^{10} \left(2 v a+r-3\right)^2}\bigg[-\frac{42 a^6}{v}+12 r (2 r-1) a^5-\frac{9 (3 r-7) a^4}{v^{3}}\nonumber\\
&+2 r^2 \left(6 r^2-20 r-1\right) a^3+\frac{2 \left(r^2+5 r-3\right) a^2}{v^{5}}+2 r^3 \left(r^3-10 r^2+21 r-12\right) a\nonumber\\
&+\frac{3 \left(r^2-3 r+2\right)}{v^{9}}+18 a^7\bigg],
\end{align}
\begin{align}
B^{\{r\}}_{\{dr \bar{m}\bar{m}\}}=&-\frac{3 \lambda }{2 r^9 \left(2 v a+r-3\right)^2}\bigg[-\frac{6 a^5}{v}+r (3 r-2) a^4-\frac{2 a^3}{v^{5}}\nonumber\\
&+r^2 \left(r^2-9 r+19\right) a^2+\frac{4 (r-3) a}{v^{5}}+r^3 \left(-2 r^2+6 r-3\right)+3 a^6\bigg],
\end{align}
\begin{align}
B^{\{\theta\}}_{\{ nn\}}=B^{\{\theta\}}_{\{ n\bar{m}\}}=B^{\{\theta\}}_{\{ \bar{m}\bar{m}\}}=B^{\{\theta\}}_{\{d\theta nn\}}=B^{\{\theta\}}_{\{d\theta n\bar{m}\}}=B^{\{\theta\}}_{\{d\theta \bar{m}\bar{m}\}}=0,
\end{align}
\begin{align}
B^{\{\phi\}}_{\{ nn\}}=&\frac{3 \lambda  v^{19} \big[(r-2) r+a^2\big] }{8 \left(2 v a+r-3\right)^2}\bigg[r (9 r-25) a^2-\frac{2 (r-5) a}{v^{3}}\nonumber\\
&+r^2 \left(3 r^2-13 r+12\right)+6 a^4\bigg],
\end{align}
\begin{align}
B^{\{\phi\}}_{\{ n\bar{m}\}}=&-\frac{3 i \lambda  }{8 \sqrt{2} r^8 \left(2 v a+r-3\right)^2}\bigg[6 v a^5+\frac{(9 r-19) a^3}{v}\nonumber\\
&+2 r (13-4 r) a^2+\frac{\left(3 r^2-5 r-10\right) a}{v^{3}}-2 (r-3) r^2-6 a^4\bigg],
\end{align}
\begin{align}
B^{\{\phi\}}_{\{ \bar{m}\bar{m}\}}=0,
\end{align}
\begin{align}
J^{\{tt\}}_{\{ nn\}}=&-\frac{3 \lambda  \big[(r-2) r+a^2\big] }{16 r^8 \left(2 v a+r-3\right)^2}\bigg[\frac{2 (r+5) a^3}{v}+2 r \left(r^2+2 r-6\right) a^2\nonumber\\
&+\frac{2 (r-3) a}{v^{5}}+r^4 (2 r-3)-a^4\bigg],
\end{align}
\begin{align}
J^{\{tt\}}_{\{ n\bar{m}\}}=&-\frac{3 i \lambda  a \left(r^2+a^2\right)}{8 \sqrt{2} r^6},
\end{align}
\begin{align}
J^{\{tt\}}_{\{ \bar{m}\bar{m}\}}=&-\frac{3 \lambda  }{8 r^6 \left(2 v a+r-3\right)^2 \left((r-2) r+a^2\right)}\bigg[-\frac{2 (r+4) a^5}{v}+r \left(-2 r^2+6 r+15\right) a^4\nonumber\\
&+\frac{4 \left(r^2-2 r-2\right) a^3}{v^{3}}+r^3 \left(-2 r^2+5 r+4\right) a^2+\frac{2 (3 r-8) a}{v^{9}}+r^5 (9-4 r)+a^6\bigg],
\end{align}
\begin{align}
J^{\{tr\}}_{\{ nn\}}=&-\frac{3 \lambda  \left(v a+r\right) }{8 r^8 \left(2 v a+r-3\right)^2}\bigg[\frac{2 a^5}{v}+r (15-8 r) a^4+\frac{4 (r-2) a^3}{v^{3}}+\frac{2 (r-2)^2 a}{v^{5}}\nonumber\\
&+r^2 \left(-7 r^2+26 r-24\right) a^2-(r-2)^2 r^3 (2 r-3)-3 a^6\bigg],
\end{align}
\begin{align}
J^{\{tr\}}_{\{ n\bar{m}\}}=&-\frac{3 i \lambda  \left(r^{3/2}+a\right) \left(v a-1\right) }{8 \sqrt{2} r^7 \big(2 v a+r-3\big)^2}\big[(r-2) r+a^2\big] \bigg[-\frac{2 a}{v}+r (2 r-3)+3 a^2\bigg],
\end{align}
\begin{align}
J^{\{tr\}}_{\{ \bar{m}\bar{m}\}}=0,
\end{align}
\begin{align}
J^{\{tr\}}_{\{dr nn\}}=&-\frac{3 \lambda  \left(r-a^2\right) \left(v a+r\right) }{4 r^9 \left(2 v a+r-3\right)^2}\big[(r-2) r+a^2\big]\bigg[\frac{2 a}{v}+r (3-2 r)-3 a^2\bigg],
\end{align}
\begin{align}
J^{\{tr\}}_{\{dr n\bar{m}\}}=&\frac{3 i \lambda  \left(r^{3/2}+a\right) \left(v a+1\right) \big[(r-2) r+a^2\big] \big[-\frac{2 a}{v}+r (2 r-3)+3 a^2\big]}{8 \sqrt{2} r^8 \left(2 v a+r-3\right)^2},
\end{align}
\begin{align}
J^{\{tr\}}_{\{dr \bar{m}\bar{m}\}}=0,
\end{align}
\begin{align}
J^{\{t\theta\}}_{\{nn\}}=0,
\end{align}
\begin{align}
J^{\{t\theta\}}_{\{n\bar{m}\}}=&-\frac{3 \lambda  \left(v a+r\right) \left((r-2) r+a^2\right) \big(\frac{4 a}{v}-r^2-3 a^2\big)}{8 \sqrt{2} r^7 \left(2 v a+r-3\right)^2},
\end{align}
\begin{align}
J^{\{t\theta\}}_{\{\bar{m}\bar{m}\}}=&-\frac{3 i \lambda  \left(r^{3/2}+a\right) \left(v a-1\right) \left(-\frac{4 a}{v}+r^2+3 a^2\right)}{4 r^6 \left(2 v a+r-3\right)^2},
\end{align}
\begin{align}
J^{\{t\theta\}}_{\{d\theta nn\}}=0,
\end{align}
\begin{align}
J^{\{t\theta\}}_{\{d\theta n\bar{m}\}}=\frac{3 i \lambda  a \left(v a+r\right) \big[(r-2) r+a^2\big] \left(\frac{4 a}{v}-r^2-3 a^2\right)}{8 \sqrt{2} r^8 \left(2 v a+r-3\right)^2},
\end{align}
\begin{align}
J^{\{t\theta\}}_{\{d\theta \bar{m}\bar{m}\}}=-\frac{3 \lambda  a \left(r^{3/2}+a\right) \left(v a-1\right) \left(-\frac{4 a}{v}+r^2+3 a^2\right)}{2 r^7 \left(2 v a+r-3\right)^2},
\end{align}
\begin{align}
J^{\{t\phi\}}_{\{nn\}}=\frac{3 \lambda  v^{15} \big[(r-2) r+a^2\big] \left[v a^3+(r-10) a^2+\frac{\left(r^2-6 r+12\right) a}{v}+r^2 (3-2 r)\right]}{8 \left(2 v a+r-3\right)^2},
\end{align}
\begin{align}
J^{\{t\phi\}}_{\{n\bar{m}\}}=-\frac{3 i \lambda  \left(r^2+2 a^2\right)}{8 \sqrt{2} r^6},
\end{align}
\begin{align}
J^{\{t\phi\}}_{\{\bar{m}\bar{m}\}}=\frac{3 \lambda  \left[(r-8) a^4\sqrt{r}-r \left(r^2-2 r+15\right) a^3+\frac{2 (r+4) a^2}{v^{3}}-r^3 \left(r^2-3 r+3\right) a+\frac{r-2}{v^{9}}-a^5\right]}{4 r^6 \left(2 v a+r-3\right)^2 \left[(r-2) r+a^2\right]},
\end{align}
\begin{align}
J^{\{rr\}}_{\{nn\}}=&\frac{3 \lambda  \left[2 a\sqrt{r}+r (3-2 r)-3 a^2\right] \left[(r-2) r+a^2\right]^3}{16 r^9 \left(2 v a+r-3\right)^2},
\end{align}
\begin{align}
J^{\{rr\}}_{\{n\bar{m}\}}=&-\frac{3 i \lambda  v^{15} \left(v a-1\right) }{8 \sqrt{2} \left(2 v a+r-3\right)^2}\bigg[\frac{2 a^5}{v}+r (15-8 r) a^4+\frac{4 (r-2) a^3}{v^{3}}\nonumber\\
&+r^2 \left(-7 r^2+26 r-24\right) a^2+\frac{2 (r-2)^2 a}{v^{5}}-(r-2)^2 r^3 (2 r-3)-3 a^6\bigg],
\end{align}
\begin{align}
J^{\{rr\}}_{\{\bar{m}\bar{m}\}}=-\frac{3 \lambda  \left(v a-1\right)^2 \left[2 a\sqrt{r}+r (3-2 r)-3 a^2\right] \left[(r-2) r+a^2\right]}{8 r^6 \left(2 v a+r-3\right)^2},
\end{align}
\begin{align}
J^{\{rr\}}_{\{drnn\}}=&\frac{3 \lambda  \left(r-a^2\right) }{4 r^{10} \left(2 v a+r-3\right)^2}\bigg[\frac{2 a^5}{v}+r (15-8 r) a^4+\frac{4 (r-2) a^3}{v^{3}}\nonumber\\
&+r^2 \left(-7 r^2+26 r-24\right) a^2+\frac{2 (r-2)^2 a}{v^{5}}-(r-2)^2 r^3 (2 r-3)-3 a^6\bigg],
\end{align}
\begin{align}
J^{\{rr\}}_{\{drn\bar{m}\}}=&\frac{3 i \lambda  \left[2 a\sqrt{r}+r (3-2 r)-3 a^2\right] \left[(r-2) r+a^2\right]}{8 \sqrt{2} r^9 \left(2 v a+r-3\right)^2}\bigg[-\frac{a^2}{v}\nonumber\\
&+(r-4) r a+\frac{1}{v^{5}}+3 a^3\bigg],
\end{align}
\begin{align}
J^{\{rr\}}_{\{dr\bar{m}\bar{m}\}}=-\frac{3 \lambda  \left(r-a^2\right) \left[2 a\sqrt{r}+r (3-2 r)-3 a^2\right] \left[(r-2) r+a^2\right]}{4 r^8 \left(2 v a+r-3\right)^2},
\end{align}
\begin{align}
J^{\{rr\}}_{\{ddrnn\}}=&-\frac{3 \lambda  \left[2 a\sqrt{r}+r (3-2 r)-3 a^2\right] \left[(r-2) r+a^2\right] }{4 r^{11} \left(2 v a+r-3\right)^2}\bigg[-3 (r-4) r a^2\nonumber\\
&+2 (r-3) r^2-5 a^4\bigg],
\end{align}
\begin{align}
J^{\{rr\}}_{\{ddrn\bar{m}\}}=\frac{3 i \lambda  a \left[-a\sqrt{r}+(r-6) r+6 a^2\right] \left[(r-2) r+a^2\right] \left[-2 a\sqrt{r}+r (2 r-3)+3 a^2\right]}{4 \sqrt{2} r^{10} \left(2 v a+r-3\right)^2},
\end{align}
\begin{align}
J^{\{rr\}}_{\{ddr\bar{m}\bar{m}\}}=-\frac{3 \lambda  \left(r+3 a^2\right) \left[2 a\sqrt{r}+r (3-2 r)-3 a^2\right] \left[(r-2) r+a^2\right]}{4 r^9 \left(2 v a+r-3\right)^2},
\end{align}
\begin{align}
J^{rab\theta}=0,
\end{align}
\begin{align}
J^{\{r\phi\}}_{\{nn\}}=&-\frac{3 \lambda  v^{17} }{8 \left(2 v a+r-3\right)^2}\bigg[2 a^5\sqrt{r}+r (15-8 r) a^4+\frac{4 (r-2) a^3}{v^{3}}+\frac{2 (r-2)^2 a}{v^{5}}\nonumber\\
&+r^2 \left(-7 r^2+26 r-24\right) a^2-(r-2)^2 r^3 (2 r-3)-3 a^6\bigg],
\end{align}
\begin{align}
J^{\{r\phi\}}_{\{n\bar{m}\}}=\frac{3 i \lambda  \left(v a-1\right) \left[2 a\sqrt{r}+r (3-2 r)-3 a^2\right] \left[(r-2) r+a^2\right]}{8 \sqrt{2} r^7 \left(2 v a+r-3\right)^2},
\end{align}
\begin{align}
J^{\{r\phi\}}_{\{\bar{m}\bar{m}\}}=0,
\end{align}
\begin{align}
J^{\{r\phi\}}_{\{drnn\}}=-\frac{3 \lambda  v^{19} \left(r-a^2\right) \left[2 a\sqrt{r}+r (3-2 r)-3 a^2\right] \left[(r-2) r+a^2\right]}{4 \left(2 v a+r-3\right)^2},
\end{align}
\begin{align}
J^{\{r\phi\}}_{\{drn\bar{m}\}}=-\frac{3 i \lambda  \left(v a+1\right) \left(2 a\sqrt{r}+r (3-2 r)-3 a^2\right) \left[(r-2) r+a^2\right]}{8 \sqrt{2} r^8 \left(2v a+r-3\right)^2},
\end{align}
\begin{align}
J^{\{r\phi\}}_{\{dr\bar{m}\bar{m}\}}=0,
\end{align}
\begin{align}
J^{\{\theta\theta\}}_{\{nn\}}=&-\frac{3 \lambda  }{16 r^9 \left(2 v a+r-3\right)^2}\bigg[\frac{4 a^5}{v}+r (12-7 r) a^4+\frac{8 (r-2) a^3}{v^{3}}\nonumber\\
&+r^2 \left(-5 r^2+16 r-12\right) a^2+\frac{4 (r-2)^2 a}{v^{5}}-(r-2)^2 r^4-3 a^6\bigg],
\end{align}
\begin{align}
J^{\{\theta\theta\}}_{\{n\bar{m}\}}=-\frac{3 i \lambda  \left(a-\sqrt{r}\right) \left[(r-2) r+a^2\right] \left[-4 a\sqrt{r}+r^2+3 a^2\right]}{8 \sqrt{2} r^8 \left(2 v a+r-3\right)^2},
\end{align}
\begin{align}
J^{\{\theta\theta\}}_{\{\bar{m}\bar{m}\}}=\frac{3 \lambda  \left(v a-1\right)^2 \left(4 a\sqrt{r}-r^2-3 a^2\right)}{8 r^6 \left(2 v a+r-3\right)^2},
\end{align}
\begin{align}
J^{\{\theta\theta\}}_{\{d\theta nn\}}=0,
\end{align}
\begin{align}
J^{\{\theta\theta\}}_{\{d\theta n\bar{m}\}}=-\frac{3 \lambda  a \left(a-\sqrt{r}\right) \left[(r-2) r+a^2\right] \left(-4 a\sqrt{r}+r^2+3 a^2\right)}{8 \sqrt{2} r^9 \left(2 v a+r-3\right)^2},
\end{align}
\begin{align}
J^{\{\theta\theta\}}_{\{d\theta \bar{m}\bar{m}\}}=-\frac{3 i \lambda  a \left(v a-1\right)^2 \left(4 a\sqrt{r}-r^2-3 a^2\right)}{4 r^7 \left(2 v a+r-3\right)^2},
\end{align}
\begin{align}
J^{\{\theta\theta\}}_{\{dd\theta nn\}}=&-\frac{3 \lambda  v^{21} a \left(v a-1\right) }{4 \left(2 v a+r-3\right)^2}\bigg[-4 a^5\sqrt{r}+r (7 r-12) a^4-\frac{8 (r-2) a^3}{v^{3}}\nonumber\\
&+r^2 \left(5 r^2-16 r+12\right) a^2-\frac{4 (r-2)^2 a}{v^{5}}+(r-2)^2 r^4+3 a^6\bigg],
\end{align}
\begin{align}
J^{\{\theta\theta\}}_{\{dd\theta n\bar{m}\}}=\frac{3 i \lambda  \left[(r-2) r+a^2\right] \left(4 a\sqrt{r}-r^2-3 a^2\right) \left[6 a^2\sqrt{r}-r (r+2) a+r^{5/2}-4 a^3\right]}{8 \sqrt{2} r^{10} \left(2 v a+r-3\right)^2},
\end{align}
\begin{align}
J^{\{\theta\theta\}}_{\{dd\theta \bar{m}\bar{m}\}}=\frac{3 \lambda  \left(v a-1\right)^2 \left(r^2+3 a^2\right) \left(-4 a\sqrt{r}+r^2+3 a^2\right)}{4 r^8 \left(2 v a+r-3\right)^2},
\end{align}
\begin{align}
J^{\{\theta\phi\}}_{\{nn\}}=0,
\end{align}
\begin{align}
J^{\{\theta\phi\}}_{\{n\bar{m}\}}=\frac{3 \lambda  \left(3 v a^2+r^{3/2}-4 a\right) \left[(r-2) r+a^2\right]}{8 \sqrt{2} r^7 \left(2 v a+r-3\right)^2},
\end{align}
\begin{align}
J^{\{\theta\phi\}}_{\{\bar{m}\bar{m}\}}=-\frac{3 i \lambda  \left(v a-1\right) \left(-4 a\sqrt{r}+r^2+3 a^2\right)}{4 r^6 \left(2 v a+r-3\right)^2},
\end{align}
\begin{align}
J^{\{\theta\phi\}}_{\{d\theta nn\}}=0,
\end{align}
\begin{align}
J^{\{\theta\phi\}}_{\{d\theta n\bar{m}\}}=-\frac{3 i \lambda  a \left(3 v a^2+r^{3/2}-4 a\right) \left[(r-2) r+a^2\right]}{8 \sqrt{2} r^8 \left(2 v a+r-3\right)^2},
\end{align}
\begin{align}
J^{\{\theta\phi\}}_{\{d\theta \bar{m}\bar{m}\}}=\frac{3 \lambda  a \left(v a-1\right) \left(4 a\sqrt{r}-r^2-3 a^2\right)}{2 r^7 \left(2 v a+r-3\right)^2},
\end{align}
\begin{align}
J^{\{\phi\phi\}}_{\{nn\}}=\frac{3 \lambda  \left[(r-2) r+a^2\right] \left[2 (2 r-5) a\sqrt{r}+r \left(r^2-8 r+12\right)+a^2\right]}{16 r^8 \left(2 v a+r-3\right)^2},
\end{align}
\begin{align}
J^{\{\phi\phi\}}_{\{n\bar{m}\}}=-\frac{3 i \lambda  a}{8 \sqrt{2} r^6},
\end{align}
\begin{align}
J^{\{\phi\phi\}}_{\{\bar{m}\bar{m}\}}=-\frac{3 \lambda  \left[4 (r-2) a^3\sqrt{r}+r \left(r^2-10 r+15\right) a^2+4 (r-2) ar^{3/2}-(r-2) r^3+a^4\right]}{8 r^6 \left(2 v a+r-3\right)^2 \left[(r-2) r+a^2\right]}.
\end{align}

\section{Overlap regime of PN and BP method}\label{ap:c}
To obtain the hybrid waveform between Post-Newtonian theory and Black hole perturbation method, we need to check the consistency within the overlap regime of these two methods.
In other words, the PN waveform taking the mass ratio expansion should agree with the BP waveform taking the PN expansion, to the relevant orders.
Technically it suffices to compare the tide-induced energy and energy flux, which we explicitly show here up to the $q^{-1}$ and $1.5$ PN order.
In order to accomplish this goal, we need to expand the components in Appendix \ref{sourcecomponents}, as well as the homogeneous solutions of the Teukolsky equation with the ingoing boundary condition for $l=2,3$ and incident amplitudes which can be found in \cite{shibata1995gravitational}.
\begin{align}
\omega R^{in}_{2m\omega}=&\frac{z^4}{30}+\frac{iz^5}{45}-\frac{11z^6}{1260}-\frac{iz^7}{420}+\frac{23z^8}{45360}+\frac{iz^9}{11340}\nonumber\\
&+\epsilon\left(-\frac{z^3}{15}-\frac{iz^4}{60}-\frac{41z^5}{3780}-\frac{31iz^6}{3780}\right)+\epsilon^2\left(\frac{z^2}{30}+\frac{iz^3}{60}\right)\\
\omega R^{in}_{3m\omega}=&\frac{z^5}{630}+\frac{iz^6}{1260}-\frac{z^7}{3780}-\frac{iz^8}{16200}+\epsilon\left(-\frac{z^4}{252}-\frac{iz^5}{756}\right)\\
B^{in}_{2m\omega}=&\frac{i}{8\omega^2}\left\{1-\epsilon\frac{\pi}{2}+i\epsilon\left(\frac{5}{4}-\gamma-\log 2\right)+O(\epsilon^2)\right\}\\
B^{in}_{3m\omega}=&-\frac{i}{8\omega^2}\left\{1-\epsilon\frac{\pi}{2}+i\epsilon\left(\frac{13}{6}-\gamma-\log 2\right)+O(\epsilon^2)\right\}
\end{align}
where $z=\omega r$ and $\epsilon=2M\omega$. With these equations and components in Appendix \ref{sourcecomponents}, we can obtain the energy flux up to the 1.5PN order from Eq.~\eqref{eq:fluxtoih}:
\begin{align}
E=\eta\,(M\Omega)^{2/3}\bigg[&-\frac{1}{2}+\frac{3}{8}(M\Omega)
^{2/3}+\frac{9}{2}\lambda(M\Omega)^{10/3}+\frac{33}{4}\lambda(M\Omega)^{4}\bigg]\\
\frac{dE}{dt}=-\frac{5}{32\eta^2\,(M\Omega)^{10/3}}\bigg[&1-\frac{1247}{336}(M\Omega)^{2/3}+18\lambda(M\Omega)^{10/3}-\frac{704}{28}\lambda(M\Omega)^{4}\bigg]
\end{align}
which are same as the corresponding PN result by keeping only the $\eta^{-1}$ order term\cite{vines2011post}. According to Eq.~\eqref{eq:orbitalphase}, we know that in the overlap regime the Post-Newtonian and Black Hole Perturbation methods are consistent.

\end{widetext}

\bibliography{refs.bib}
\end{document}